\documentclass[journal]{IEEEtran}
\IEEEoverridecommandlockouts

\usepackage{amsmath,amssymb,amsfonts}
\usepackage{array}
\usepackage[caption=false,font=normalsize,labelfont=sf,textfont=sf]{subfig}
\usepackage{textcomp}
\usepackage{stfloats}
\usepackage{verbatim}
\usepackage{graphicx}
\usepackage{cite}
\usepackage{tabularx}
\usepackage{booktabs}
\usepackage{enumitem}
\usepackage{tikz}
\usetikzlibrary{arrows.meta,positioning,calc,backgrounds,shapes.geometric,decorations.markings,decorations.pathreplacing}
\usepackage{circuitikz}
\usepackage{pgfplots}
\pgfplotsset{compat=1.18}
\pgfplotsset{golden axis/.style={scale only axis,width=0.8232\columnwidth,height=0.8232\columnwidth/1.618034}}
\definecolor{pltred}{RGB}{214,39,40}
\definecolor{pltgreen}{RGB}{0,158,115}
\definecolor{pltblue}{RGB}{0,114,178}
\definecolor{pltorange}{RGB}{230,159,0}
\definecolor{pltpurple}{RGB}{148,103,189}
\definecolor{pltgray}{RGB}{105,105,105}
\pgfplotsset{curve/.style={color=#1,line width=1pt,mark options={solid,fill=#1,draw=black!75,line width=0.4pt}}}
\usepackage{algorithm}
\usepackage{algpseudocode}
\algrenewcommand\algorithmicrequire{\textbf{Input:}}
\algrenewcommand\algorithmicensure{\textbf{Output:}}

\def\BibTeX{{\rm B\kern-.05em{\sc i\kern-.025em b}\kern-.08em
    T\kern-.1667em\lower.7ex\hbox{E}\kern-.125emX}}

\usepackage{acronym}
\acrodef{5G}{the fifth generation}
\acrodef{6G}{the sixth generation}
\acrodef{MIMO}{multiple-input multiple-output}
\acrodef{MISO}{multiple-input single-output}
\acrodef{RF}{radio frequency}
\acrodef{BS}{base station}
\acrodef{DAC}{digital-to-analog converter}
\acrodef{ADC}{analog-to-digital converter}
\acrodef{DSP}{digital signal processing}
\acrodef{AWGN}{additive white Gaussian noise}
\acrodef{MiLAC}{microwave linear analog computer}
\acrodef{RIS}{reconfigurable intelligent surface}
\acrodef{BD-RIS}{beyond-diagonal reconfigurable intelligent surface}
\acrodef{CSI}{channel state information}
\acrodef{ZF}{zero-forcing}
\acrodef{ZFBF}{zero-forcing beamforming}
\acrodef{SVD}{singular value decomposition}
\acrodef{PSD}{positive semidefinite}
\acrodef{PD}{positive definite}
\acrodef{MSE}{mean squared error}
\acrodef{MMSE}{minimum mean squared error}
\acrodef{LMMSE}{linear minimum mean square error}
\acrodef{RZF}{regularized zero-forcing}
\acrodef{RLS}{regularized least squares}
\acrodef{LS}{least squares}
\acrodef{MF}{matched filter}
\acrodef{GLS}{generalized least squares}
\acrodef{GMF}{generalized matched filtering}
\acrodef{OLS}{ordinary least squares}
\acrodef{OMF}{ordinary matched filtering}
\acrodef{FC}{fully-connected}
\acrodef{SNR}{signal-to-noise ratio}
\acrodef{SER}{symbol error rate}
\acrodef{QAM}{quadrature amplitude modulation}
\acrodef{NMSE}{normalized mean squared error}
\acrodef{SINR}{signal-to-interference-plus-noise ratio}
\acrodef{i.i.d.}{independent and identically distributed}
\usepackage{amsthm}
\newtheoremstyle{zackplain}%
  {\topsep}{\topsep}{\itshape}{}{\bfseries}{:}{.5em}{}
\theoremstyle{zackplain}
\newtheorem{lemma}{\textbf{Lemma}}
\newtheorem{theorem}{\textbf{Theorem}}
\newtheorem{proposition}{\textbf{Proposition}}
\newtheorem{corollary}{\textbf{Corollary}}

\newtheorem{remark}{\textbf{Remark}}
\renewenvironment{proof}{\par\noindent\textit{\textbf{Proof:}}\ }{\hfill$\square$\par\medskip}
\usepackage[bottom=1.02in,top=0.75in,left=0.625in,right=0.625in]{geometry}
\usepackage{units}
\usepackage{bm}
\newcommand{\TT}{\mathsf{T}}
\newcommand{\HH}{\mathsf{H}}

\newcommand{\realp}[1]{\operatorname{Re}\!\left\{#1\right\}}
\newcommand{\imagp}[1]{\operatorname{Im}\!\left\{#1\right\}}

\newcommand{\iv}{{\bf i}}

\newcommand{\nv}{{\bf n}}

\newcommand{\uv}{{\bf u}}

\newcommand{\vv}{{\bf v}}
\newcommand{\xv}{{\bf x}}
\newcommand{\yv}{{\bf y}}
\newcommand{\zv}{{\bf z}}

\newcommand{\Am}{{\bf A}}
\newcommand{\Bm}{{\bf B}}
\newcommand{\Cm}{{\bf C}}
\newcommand{\Dm}{{\bf D}}
\newcommand{\Em}{{\bf E}}
\newcommand{\Fm}{{\bf F}}
\newcommand{\Gm}{{\bf G}}
\newcommand{\Hm}{{\bf H}}
\newcommand{\Id}{{\bf I}}

\newcommand{\Km}{{\bf K}}

\newcommand{\Pm}{{\bf P}}
\newcommand{\Qm}{{\bf Q}}

\newcommand{\Sm}{{\bf S}}

\newcommand{\Um}{{\bf U}}
\newcommand{\Wm}{{\bf W}}
\newcommand{\Vm}{{\bf V}}

\newcommand{\Ym}{{\bf Y}}

\newcommand{\Zerom}{{\bf 0}}

\newcommand{\Gammam}{\hbox{\boldmath$\Gamma$}}

\newcommand{\Phim}{\hbox{\boldmath$\Phi$}}

\usepackage{amsmath}

\DeclareMathOperator{\diag}{diag}
\DeclareMathOperator{\rank}{rank}
\newcommand{\lift}[1]{\Psi(#1)}          
\newcommand{\lmin}{\lambda_{\min}}
\newcommand{\lmax}{\lambda_{\max}}
\definecolor{cport}{RGB}{30,64,124}
\definecolor{ccoup}{RGB}{20,115,85}
\definecolor{cshunt}{RGB}{200,90,15}
\definecolor{csrc}{RGB}{110,35,120}
\definecolor{cread}{RGB}{165,25,45}
\definecolor{caux}{RGB}{90,90,95}

\begin{document}

\title{Can Lossless and Reciprocal Microwave Linear Analog Computers (MiLACs) Compute the LMMSE Estimator With Low Computational Complexity?}

\author{Yuchen Zhang, \emph{Member, IEEE}, Bruno Clerckx, \emph{Fellow, IEEE}, and Tareq Y. Al-Naffouri, \emph{Fellow, IEEE}
\thanks{Y. Zhang and T. Y. Al-Naffouri are with the Electrical and Computer Engineering Program, Computer, Electrical and Mathematical Sciences and Engineering (CEMSE), King Abdullah University of Science and Technology (KAUST), Thuwal 23955-6900, Kingdom of Saudi Arabia (e-mail: \{yuchen.zhang, tareq.alnaffouri\}@kaust.edu.sa).}
\thanks{B. Clerckx is with the Department of Electrical and Electronic Engineering, Imperial College London, London, SW7 2AZ, U.K. (email: b.clerckx@imperial.ac.uk).}
}

\maketitle

\begin{abstract}
A microwave linear analog computer (MiLAC) is a tunable microwave network that performs linear algebraic operations as signals propagate through it. Uniquely, a MiLAC can compute common signal processing functions, such as the linear minimum mean square error (LMMSE) estimate, entirely in the analog domain, without digital operations or matrix inversion during estimation. Placed between the antennas and the radio frequency (RF) chains of a receiver, a MiLAC needs digital operations only to reconfigure its admittance matrix. When the channel changes, the number of these operations grows linearly with the number of antennas, whereas computing the LMMSE estimate digitally incurs quadratic complexity for a general noise covariance. In hardware, a MiLAC is preferably lossless and reciprocal, so that it can be built from tunable capacitors and inductors only, which reduces power consumption and simplifies implementation. However, the existing MiLAC design for LMMSE estimation requires a complex Hermitian admittance matrix, whereas a lossless and reciprocal MiLAC must have a purely imaginary and symmetric one. In this paper, we use two techniques to design lossless and reciprocal MiLACs that compute the LMMSE estimate exactly, up to a known scaling factor. First, doubling the input and output ports embeds the complex channel in a purely imaginary and symmetric admittance matrix, which yields the regularized least squares (RLS) estimate. Second, the general LMMSE estimate also requires real covariance-dependent diagonal blocks in the admittance matrix, which a purely imaginary matrix cannot provide. We therefore couple the input or output ports to auxiliary ports, which are neither driven nor read and whose terminations provide the covariance-dependent weighting, while the MiLAC itself remains lossless and reciprocal. The proposed MiLAC retains the linear computational complexity of the existing design. Moreover, its number of tunable components grows linearly with the number of antennas for white noise, with or without interference from a few sources, rather than quadratically as for the fully-connected existing design.
\end{abstract}


\section{Introduction}
\label{sec:intro}

Digital computing is versatile, precise, and robust to noise, but it faces two limitations \cite{Nerini2025AnalogI}. First, its power consumption is high, especially because of power-hungry \acp{ADC} and \acp{DAC}. Second, its speed is limited by the processor clock and by the sampling rate of the converters \cite{Nerini2025AnalogI}. These limitations have revived the interest in analog computing, which processes continuous-valued signals directly as they arrive \cite{Caloz2013Analog} and offers energy efficiency and massive parallelism \cite{Nerini2025AnalogI}.

Modern analog computers perform linear algebra through the physical response of a device rather than with a clocked processor. Optical systems can serve as analog accelerators for electronic computers \cite{Solli2015Optical}, including for matrix inversion \cite{Wu2014Optical}, and metamaterials and metasurfaces can shape an incident wave to perform mathematical operations and to solve equations \cite{Silva2014Performing,Estakhri2019Inverse}. Similarly, resistive memory arrays compute a matrix-vector product, or solve a linear system, in one step \cite{Sun2019Solving}, and have been used for wireless signal processing \cite{Wang2023Parallel}. All these efforts share the objective of replacing an expensive digital computation, matrix inversion in particular, with analog hardware.

Recently, the \ac{MiLAC} has been proposed in \cite{Nerini2025AnalogI,Nerini2025AnalogII} to pursue this objective at microwave frequencies. A \ac{MiLAC} is a multiport microwave network of tunable admittance components. Signals applied at its input ports propagate through the network and are read, linearly transformed, at its output ports \cite{Nerini2025AnalogI}. Specifically, the tunable components set the admittance matrix of the network, and the port voltages solve a linear system defined by this matrix, so the network computes matrix inversions and matrix products with no digital operation \cite{Nerini2025AnalogI}. Designing a \ac{MiLAC} for a given task therefore amounts to choosing its admittance matrix. In particular, inverting a matrix only requires configuring the network, with a number of real operations that grows quadratically with the matrix size, rather than cubically as for a digital inversion \cite{Nerini2025AnalogI}.

Since the \ac{LMMSE} estimator involves a matrix inversion, a \ac{MiLAC} can compute it, together with its special case, the \ac{RLS} estimator \cite{Nerini2025AnalogI,Nerini2025AnalogII}. At a receiver, the \ac{MiLAC} computes the \ac{LMMSE} estimate entirely in the analog domain, and digital operations are needed only to set its tunable components \cite{Nerini2025AnalogI}. When the channel changes, the number of these operations grows linearly with the number of antennas, whereas computing the \ac{LMMSE} estimate digitally with a general noise covariance requires a number that grows quadratically \cite{Nerini2025AnalogI}.

Beyond these computational capabilities, \acp{MiLAC} are particularly attractive for very large antenna arrays \cite{Marzetta2010Noncooperative,Bjornson2024Gigantic}. With such arrays, a digital implementation is dominated by the cost of the \ac{RF} chains, of their high-resolution \acp{ADC} and \acp{DAC}, and of the per-symbol digital processing \cite{Nerini2025AnalogII}. Since a \ac{MiLAC} processes the microwave signals directly, it can be placed between the antennas and the \ac{RF} chains to perform the precoding at a transmitter, or the combining at a receiver, in the analog domain. This reduces all three costs and enables gigantic \ac{MIMO} beamforming \cite{Nerini2025AnalogII,Wei2026Gigantic}. In particular, with one \ac{RF} chain per stream and arbitrary tunable admittances, \ac{MiLAC}-aided beamforming retains the flexibility of fully digital beamforming \cite{Nerini2025AnalogII}. Conventional hybrid beamforming, in contrast, is guaranteed to do so with twice as many \ac{RF} chains as streams \cite{Sohrabi2016Hybrid}. Following \cite{Nerini2025AnalogI,Nerini2025AnalogII}, \acp{MiLAC} have been studied for capacity-achieving \ac{MIMO} \cite{Nerini2025Capacity} with reduced circuit complexity \cite{Nerini2026Stem}, multiuser beamforming \cite{Wu2026MiLAC,Zhang2026Beamforming,Fang2026Performance,Zhou2026TwoLayer}, simultaneous active and passive beamforming \cite{Nerini2026ActivePassive}, spectral and energy efficiency \cite{Zhang2026EE,Zhou2026LossyEE}, channel and direction-of-arrival estimation \cite{Zhang2026ChannelEst,Zhang2026ChannelBF,Zhang2026DOA,Liu2026Radar}, over-the-air computation \cite{Wang2026AirComp}, and integrated sensing and communication \cite{Zhang2026ISAC}, with physics-compliant modeling \cite{Nerini2026Physics} and a hardware prototype \cite{Nerini2026Hybrid}.

The early works \cite{Nerini2025AnalogI,Nerini2025AnalogII} focus on tunable admittances that can take any complex value. In contrast, most of the subsequent works consider a lossless and reciprocal \ac{MiLAC} \cite{Nerini2025Capacity,Nerini2026Stem,Wu2026MiLAC,Zhang2026Beamforming,Fang2026Performance,Zhou2026TwoLayer,Nerini2026ActivePassive,Zhang2026EE,Zhang2026DOA,Liu2026Radar,Wang2026AirComp,Zhang2026ISAC,Nerini2026Physics,Nerini2026Hybrid}. A lossless \ac{MiLAC} neither dissipates nor generates signal power, and a reciprocal \ac{MiLAC} transmits identically in both directions between any two of its ports \cite{Wu2026MiLAC}. Accordingly, a lossless \ac{MiLAC} needs no costly, power-consuming active components \cite{Nerini2025AnalogI}, and a reciprocal one needs no non-reciprocal devices, which facilitates its implementation \cite{Wu2026MiLAC}. Although neither property holds exactly in practice, both are standard idealizations of passive hardware built from tunable capacitors and inductors. Together, however, the two properties restrict the admittance matrix of the \ac{MiLAC} to be purely imaginary and symmetric \cite{Nerini2025Capacity}, a strict subset of the matrices realizable with arbitrary tunable admittances.

Under this restriction, the existing \ac{MiLAC} computation of the \ac{RLS} and \ac{LMMSE} estimators no longer applies. Specifically, the construction of \cite{Nerini2025AnalogI} sets the admittance matrix to a complex Hermitian matrix assembled from the channel and the covariances, so that the estimator appears as a block of the matrix inverse that the \ac{MiLAC} computes. The real part of this admittance matrix corresponds to conductances, which dissipate power where positive and require active components where negative. Moreover, unless the channel and the covariances are real, this Hermitian matrix is not symmetric, which requires non-reciprocal components. Realizing it thus requires a \ac{MiLAC} whose admittance matrix can take any complex value, which we call \emph{unconstrained}, as in \cite{Wu2026MiLAC}. A lossless and reciprocal network could still realize a scaled estimator, since any matrix of spectral norm at most one is a block of the scattering matrix of such a network \cite{Wu2026MiLAC}. However, the construction in \cite{Wu2026MiLAC} starts from the matrix itself and its singular value decomposition, so the estimator, matrix inversion included, would first have to be computed digitally. This removes the computational advantage of the \ac{MiLAC}. \emph{Whether a lossless and reciprocal \ac{MiLAC} can compute the \ac{LMMSE} estimator for arbitrary noise and prior covariances, with no digital matrix inversion when the channel changes, has thus remained open.}

Besides the computational complexity, the circuit complexity of a \ac{MiLAC}, defined as its number of tunable components \cite{Nerini2026Stem}, becomes critical for large arrays. The \ac{MiLAC} that computes the \ac{LMMSE} estimator in \cite{Nerini2025AnalogI} connects every port to ground and to every other port, so its number of components grows quadratically with the number of antennas \cite{Nerini2025AnalogI}. To reduce this number, stem-connected \acp{MiLAC}, in which only a few central ports are connected to all other ports, have been proposed in \cite{Nerini2026Stem}. Their number of components grows linearly with the number of antennas \cite{Nerini2026Stem,Zhang2026Beamforming}. However, a stem-connected \ac{MiLAC} is configured to realize a prescribed matrix computed digitally, such as the singular vectors of the channel in \cite{Nerini2026Stem}. Computing the \ac{LMMSE} estimator with such a \ac{MiLAC} would thus require digital computation, matrix inversion included, as for the construction of \cite{Wu2026MiLAC}. To date, no \ac{MiLAC} computes the \ac{LMMSE} estimator in a lossless and reciprocal network, with no digital matrix inversion, and with a number of components that grows linearly with the number of antennas.

In this paper, we address this problem for receive combining by explicit construction, and leave transmit beamforming to future work. The proposed \acp{MiLAC} compute the \ac{LMMSE} estimator for arbitrary noise and prior covariances, and, for white noise with or without interference from a few sources, their number of components grows only linearly with the number of antennas. Our construction uses two techniques. First, we double the input and output ports and represent each complex matrix by a real matrix of twice the size. The doubled ports embed the complex channel in a purely imaginary and symmetric admittance matrix, and the \ac{MiLAC} then computes the \ac{RLS} estimator, which involves no covariance matrix. One $90^\circ$ hybrid coupler per antenna feeds a pair of input ports, and one per stream combines a pair of output ports into one \ac{RF} chain. Second, the \ac{LMMSE} estimator also requires real covariance-dependent matrices in the diagonal blocks of the admittance matrix, which a purely imaginary matrix cannot provide, even with doubled ports. We therefore add \emph{auxiliary ports}, neither driven nor connected to an \ac{RF} chain, and couple them to the input or output ports. Through this coupling, their resistive terminations provide the covariance-dependent weighting, while the \ac{MiLAC} itself remains lossless and reciprocal. This linear growth is due to the sparsity of the resulting network: the channel couples the input ports only to the output ports, and, for such noise, the number of auxiliary ports does not grow with the number of antennas. The main contributions of this paper are summarized as follows.
\begin{itemize}
\item \emph{Lossless and reciprocal realization of the estimators:} For the first time, we construct lossless and reciprocal \acp{MiLAC} that compute the \ac{RLS} and \ac{LMMSE} estimators \emph{exactly}, up to a known scaling factor. Every component value is given \emph{in closed form} by the channel and by factors computed offline from the covariances. Each \ac{RF} chain receives the estimate of one stream up to this factor, which a fixed gain removes, so one \ac{RF} chain per stream suffices.
\item \emph{Linear computational complexity:} It was shown in \cite{Nerini2025AnalogI} that an unconstrained \ac{MiLAC} computes the \ac{LMMSE} estimate with a number of real operations per coherence block that grows linearly with the number of antennas, whereas a digital receiver with a general noise covariance needs a number that grows quadratically. We show that a lossless and reciprocal \ac{MiLAC} \emph{retains this linear complexity}, with the same leading term as in \cite{Nerini2025AnalogI}. Specifically, the quantities that depend only on the covariances, which involve matrix inversion and factorization, are computed once offline. When the channel changes, only the channel-dependent component values are rewritten, and the \ac{MiLAC} itself forms the estimate with no per-symbol arithmetic.
\item \emph{Linear circuit complexity:} We derive the number of tunable components of the proposed realizations and show that it grows only \emph{linearly} with the number of antennas, for \ac{RLS} and for \ac{LMMSE} with white noise, with or without interference from a few sources, whereas that of the fully-connected \ac{MiLAC} of \cite{Nerini2025AnalogI} grows quadratically. Notably, the growth remains linear even though such interference makes the noise covariance a dense matrix, since a few auxiliary ports suffice for the covariance-dependent weighting. The proposed \acp{MiLAC} combine two benefits: they compute the \ac{LMMSE} estimator in a lossless and reciprocal network and, for such noise, with a number of components that grows linearly.
\item \emph{Numerical validation:} In a simulated uplink with $8$ users and $256$ antennas, the \ac{MiLAC} attains the digital \ac{LMMSE} sum rate with up to $466$ times fewer real operations.
\end{itemize}

\subsection{Paper Organization and Notation}

The rest of this paper is organized as follows. Section~\ref{sec:model} introduces the system model and the lossless and reciprocal class, and states the problem. Section~\ref{sec:rzf} realizes the \ac{RLS} estimator by doubling the ports, and Section~\ref{sec:lmmse} extends this realization to the \ac{LMMSE} estimator with auxiliary ports. Section~\ref{sec:hw} analyzes the circuit complexity of both realizations, Section~\ref{sec:sim} presents numerical results, and Section~\ref{sec:conc} concludes the paper. The Appendix collects the properties of the real-valued representation and all proofs.

\emph{Notation:} Vectors and matrices are written in bold lowercase and bold uppercase, scalars in italic, and $j=\sqrt{-1}$ is the imaginary unit. Transpose, conjugate transpose, complex conjugate, and inverse are $(\cdot)^\TT$, $(\cdot)^\HH$, $(\cdot)^\star$, and $(\cdot)^{-1}$. The spectral norm is $\|\cdot\|_2$, the Frobenius norm is $\|\cdot\|_F$, the rank is $\rank(\cdot)$, the smallest and largest eigenvalues of a Hermitian matrix are $\lmin(\cdot)$ and $\lmax(\cdot)$, and $\Am\succ\Zerom$ means that $\Am$ is Hermitian positive definite. We write $\realp{\cdot}$ and $\imagp{\cdot}$ for the real and imaginary parts, $[\Am]_{i,k}$ for the $(i,k)$th entry of $\Am$, $\Id_n$ and $\Zerom_n$ for the $n\times n$ identity and zero matrices, and $\Id$ and $\Zerom$ for an identity matrix and a zero vector or matrix whose dimensions are clear from the context. The symbol $\triangleq$ denotes a definition.

\section{System Model and Problem Statement}
\label{sec:model}

In this section, we first model a \ac{MiLAC} with multiport circuit theory and recall how prior work computes estimators with \acp{MiLAC}. We then define the lossless and reciprocal class and state the problem addressed in this paper.

\subsection{Multiport Networks and the Admittance Matrix}
\label{sec:multiport}

We consider narrowband signals, modeled as sinusoids at the carrier frequency. Each signal is represented by a complex \emph{phasor}, whose magnitude and angle are the amplitude and the phase of the sinusoid relative to a common reference \cite{Nerini2025AnalogI}. A linear microwave network is accessed through external terminals called \emph{ports}. Let $\vv=[v_1,\dots,v_P]^\TT$ collect the voltage phasors at the $P$ ports of a network and $\iv=[i_1,\dots,i_P]^\TT$ the current phasors, with each current entering the network. By linearity, the two are related through the \emph{admittance matrix} $\Ym\in\mathbb{C}^{P\times P}$ as
\begin{equation}
\label{eq:admittance}
\iv=\Ym\vv,
\end{equation}
which characterizes the network independently of its internal circuit topology \cite{Nerini2025AnalogI}.

A \ac{MiLAC} realizes a desired $\Ym$ with \emph{tunable admittance components}. The real part of an admittance is its conductance, and its imaginary part is its susceptance, which is realized by a tunable capacitor when positive and by a tunable inductor when negative. We denote by $Y_{k,k}$ the \emph{port-to-ground} component, which connects port $k$ to ground, and by $Y_{i,k}$, $i\ne k$, the \emph{inter-port} component between ports $i$ and $k$ that determines $[\Ym]_{i,k}$, with $Y_{i,k}=0$ where no component is present. The entries of the admittance matrix follow from these components as \cite{Nerini2025AnalogI}
\begin{equation}
\label{eq:stamp}
[\Ym]_{i,k}=
\begin{cases}
-\,Y_{i,k}, & i\ne k,\\[2pt]
\displaystyle\sum_{p=1}^{P}Y_{p,k}, & i=k,
\end{cases}
\end{equation}
that is, an off-diagonal entry is minus the inter-port component between the two ports, and a diagonal entry is the sum of all components connected to the port. Conversely, inverting \eqref{eq:stamp} gives each component from the matrix \cite{Nerini2025AnalogI},
\begin{equation}
\label{eq:components}
\underbrace{Y_{i,k}=-[\Ym]_{i,k}}_{\text{Inter-port }(i\ne k)},\qquad
\underbrace{Y_{k,k}=\sum_{p=1}^{P}[\Ym]_{p,k}}_{\text{Port-to-ground}},
\end{equation}
so a port-to-ground component is a column sum of $\Ym$, which is generally nonzero even where the diagonal entry $[\Ym]_{k,k}$ is zero. When $Y_{i,k}$ and $Y_{k,i}$ are tuned independently, which requires non-reciprocal components, these $P^2$ components can realize any $\Ym\in\mathbb{C}^{P\times P}$ \cite{Nerini2025AnalogI}. In contrast, an ordinary two-terminal component, such as a capacitor or an inductor, gives $Y_{i,k}=Y_{k,i}$, so a network of such components has a symmetric admittance matrix by \eqref{eq:stamp} \cite{Nerini2025Capacity}.

\subsection{The MiLAC and Its Input--Output Relationship}
\label{sec:io}

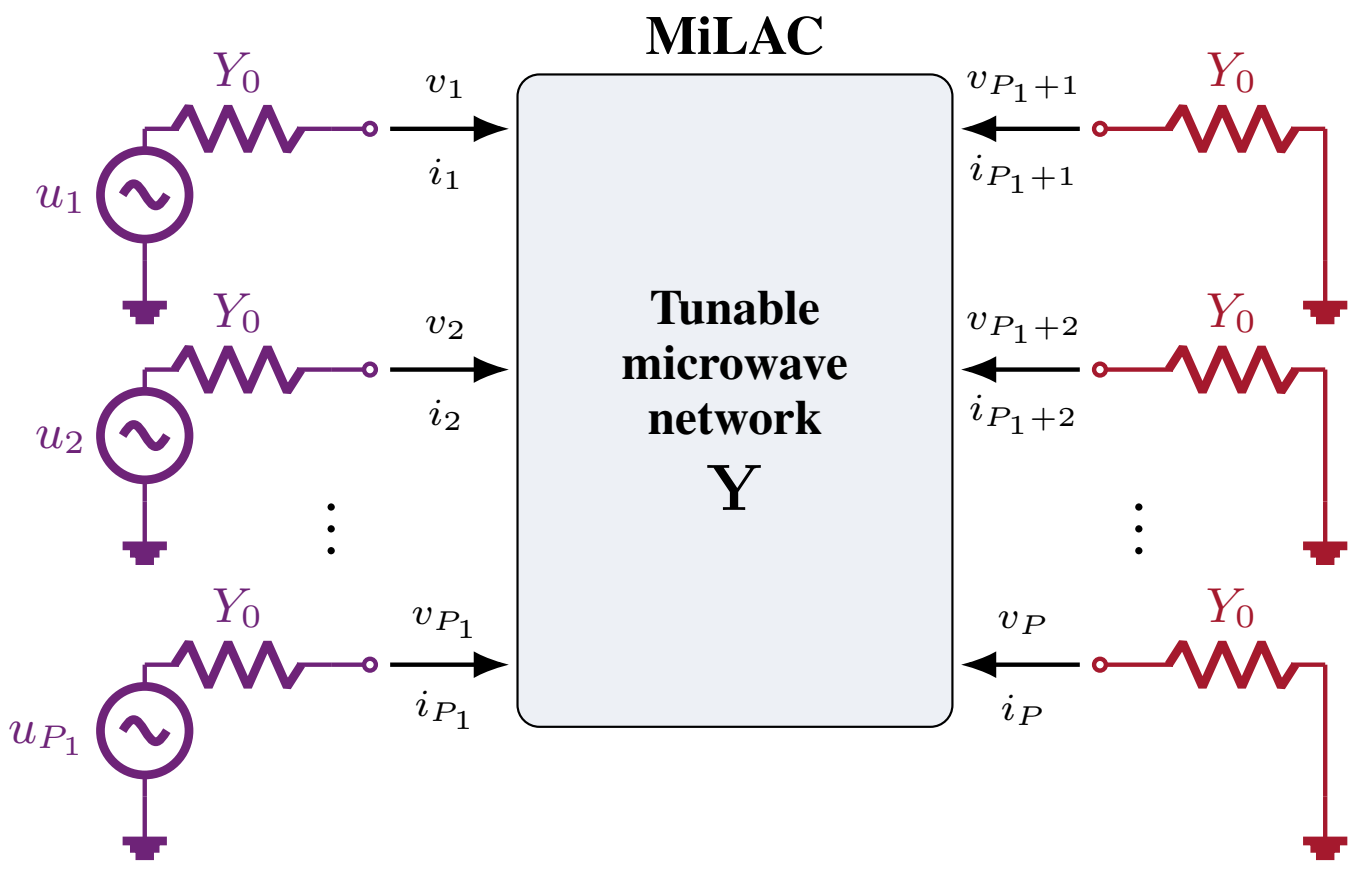
\begin{figure}[t]
\centering
\resizebox{\columnwidth}{!}{%
\ctikzset{bipoles/length=0.95cm}%
\begin{circuitikz}[transform shape, >=Latex, every node/.style={font=\small}]
\node[draw,rounded corners,fill=cport!8,minimum width=2.8cm,minimum height=4.2cm,align=center,font=\bfseries\small] (net) at (0,0){Tunable\\ microwave\\ network\\[4pt] \large $\Ym$};
\node[font=\bfseries] at (0,2.35){MiLAC};
\foreach \y/\lab in {1.75/1, 0.2/2, -1.7/{P_1}}{
  \draw[csrc,thick] (-3.8,\y) to[vsourcesin,l_=$u_{\lab}$] (-3.8,\y-0.85) node[ground]{};
  \draw[csrc,thick] (-3.8,\y) to[R,l=$Y_0$] (-2.6,\y);
  \draw[csrc,thick] (-2.6,\y) to[short,-o] (-2.35,\y);
  \draw[->,thick] (-2.22,\y) -- (-1.45,\y);
  \node[font=\scriptsize] at (-1.86,\y+0.27){$v_{\lab}$};
  \node[font=\scriptsize] at (-1.86,\y-0.29){$i_{\lab}$};
}
\node[font=\large] at (-2.6,-0.72){$\vdots$};
\foreach \y/\lab in {1.75/{P_1{+}1}, 0.2/{P_1{+}2}, -1.7/{P}}{
  \draw[->,thick] (2.22,\y) -- (1.45,\y);
  \node[font=\scriptsize] at (1.86,\y+0.27){$v_{\lab}$};
  \node[font=\scriptsize] at (1.86,\y-0.29){$i_{\lab}$};
  \draw[cread,thick] (2.35,\y) to[short,o-] (2.6,\y) to[R,l=$Y_0$] (3.8,\y);
  \draw[cread,thick] (3.8,\y) -- (3.8,\y-0.85) node[ground]{};
}
\node[font=\large] at (2.6,-0.72){$\vdots$};
\end{circuitikz}}
\caption{A \ac{MiLAC} with $P=P_1+P_2$ ports and admittance matrix $\Ym$. The $P_1$ input ports are driven by the voltage sources $u_1,\dots,u_{P_1}$ in series with the reference admittance $Y_0$, and the $P_2$ output ports are terminated in $Y_0$, where the voltages $v_{P_1+1},\dots,v_{P}$ are read.}
\label{fig:model}
\end{figure}

Fig.~\ref{fig:model} shows a \ac{MiLAC} with $P$ ports, of which $P_1$ are \emph{input} ports and $P_2=P-P_1$ are \emph{output} ports. Input port $p$ is driven by an ideal voltage source of voltage $u_p$ in series with the reference admittance $Y_0$, e.g., $Y_0=1/(50\,\Omega)$, and the source voltages are collected in $\uv=[u_1,\dots,u_{P_1}]^\TT$. Each output port is terminated in $Y_0$, and its voltage is the output that is read \cite{Nerini2025AnalogI}. Denote by $\vv_1,\iv_1\in\mathbb{C}^{P_1}$ the voltages and currents at the input ports and by $\vv_2,\iv_2\in\mathbb{C}^{P_2}$ those at the output ports. Applying Ohm's law at the sources and at the terminations gives \cite{Nerini2025AnalogI}
\begin{equation}
\label{eq:interface}
\iv_1=Y_0(\uv-\vv_1)\ \ \text{(Input)},\qquad \iv_2=-Y_0\vv_2\ \ \text{(Output)}.
\end{equation}
Let $\tilde\uv=[\uv^\TT,\,\Zerom^\TT]^\TT$ denote $\uv$ padded with zeros at the output ports. Combining these relations with \eqref{eq:admittance} yields the input--output relationship \cite{Nerini2025AnalogI}
\begin{equation}
\label{eq:iomap}
\vv=\Pm^{-1}\tilde\uv,\qquad \Pm\triangleq\frac{\Ym}{Y_0}+\Id_P ,
\end{equation}
which holds when $\Pm$ is invertible. As a result, the voltages at the $P_2$ output ports are $\vv_2=\Wm\uv$, where $\Wm$ is the \emph{output-by-input block} of $\Pm^{-1}$,
\begin{equation}
\label{eq:Wblock}
\Wm=\bigl[\Pm^{-1}\bigr]_{21}\in\mathbb{C}^{P_2\times P_1}.
\end{equation}
The \ac{MiLAC} thus computes a block of a matrix inverse, set by its tunable components through \eqref{eq:stamp}, as the applied signals propagate through the network. Designing a \ac{MiLAC} amounts to choosing $\Ym$ so that $\Wm$ is the desired matrix.

\subsection{Computing Estimators With a MiLAC}
\label{sec:place}

The linear estimators that we consider at a receiver also involve a matrix inverse. Consider a vector $\xv\in\mathbb{C}^M$ of $M$ transmitted streams that is observed through a channel $\Hm\in\mathbb{C}^{N\times M}$ at $N$ antennas as $\yv=\Hm\xv+\nv\in\mathbb{C}^N$, where $\nv$ is the noise, and $M\ll N$ in the large arrays of interest. A linear estimator forms the estimate $\hat\xv$ of $\xv$ by applying a matrix to $\yv$. Two widely used linear estimators in wireless are
\begin{align}
\label{eq:rzf}
\Wm_{\mathrm{RLS}}&=(\Hm^\HH\Hm+\lambda\Id_M)^{-1}\Hm^\HH,\\
\label{eq:lmmse}
\Wm_{\mathrm{LMMSE}}&=(\Hm^\HH\Cm_n^{-1}\Hm+\Cm_x^{-1})^{-1}\Hm^\HH\Cm_n^{-1},
\end{align}
where $\lambda>0$ is a regularization weight that trades interference suppression against noise amplification, $\Cm_n\succ\Zerom$ is the noise covariance, and $\Cm_x\succ\Zerom$ is the prior covariance of $\xv$, with $\xv$ and $\nv$ zero-mean and uncorrelated. Equation~\eqref{eq:lmmse} is the \ac{LMMSE} estimator \cite{Nerini2025AnalogI,Kay1993Fundamentals}, and \eqref{eq:rzf} is its special case for white noise and an isotropic prior, $\Cm_n=\Id$ and $\Cm_x=\lambda^{-1}\Id$, known as the \ac{RLS} estimator or ridge regression \cite{Nerini2025AnalogII}. As $\lambda\to0$, the \ac{RLS} estimator tends to the least-squares estimator, which is the \ac{ZF} estimator when $\Hm$ has full column rank, and for large $\lambda$ it becomes proportional to the \ac{MF}. In transmit beamforming, the same regularized inverse gives the regularized channel inversion precoder \cite{Peel2005Vector}, known as \ac{RZF} beamforming \cite{Nerini2025AnalogII}. Both estimators invert a Hermitian positive definite $\Km$, namely $\Hm^\HH\Hm+\lambda\Id_M$ for \ac{RLS} and $\Hm^\HH\Cm_n^{-1}\Hm+\Cm_x^{-1}$ for \ac{LMMSE}.

Since each estimator can be written as a block of a matrix inverse, like the \ac{MiLAC} output \eqref{eq:Wblock}, prior work computes these estimators with a \ac{MiLAC} by matching the two. Specifically, if the four blocks of $\Pm$ are set to the channel and the covariances as
\begin{equation}
\label{eq:hermplace}
\Pm=\begin{bmatrix}\Cm_n & \Hm\\[1pt] \Hm^\HH & -\Cm_x^{-1}\end{bmatrix},
\end{equation}
block matrix inversion shows that the output-by-input block of $\Pm^{-1}$ is $\Wm_{\mathrm{LMMSE}}$ \cite{Nerini2025AnalogI}, and the case $\Cm_n=\Id$, $\Cm_x=\lambda^{-1}\Id$ gives $\Wm_{\mathrm{RLS}}$. The estimator is thus obtained by setting the components so that $\Pm$ equals \eqref{eq:hermplace}, applying $\yv$ at the input ports, and reading $\hat\xv$ at the output ports. Whereas a digital receiver must form and invert the $M\times M$ matrix $\Km$ whenever the channel changes, the estimate here appears at the output ports as the signals propagate, \emph{without digital matrix inversion}. This is the computational advantage of a \ac{MiLAC} for estimation.

\subsection{The Lossless and Reciprocal Class}
\label{sec:lr}

The construction of prior work, however, requires a \ac{MiLAC} that can realize \eqref{eq:hermplace}, which a lossless and reciprocal \ac{MiLAC} cannot. A network built only from tunable susceptances, which neither dissipate nor generate power and respond identically in both directions, is \emph{lossless} and \emph{reciprocal}. The following proposition, standard in microwave network theory \cite{Pozar2011Microwave}, gives the admittance matrices that these two properties allow \cite{Nerini2025Capacity}.

\begin{proposition}[Lossless and reciprocal class]
\label{prop:class}
A multiport network is lossless and reciprocal if and only if $\Ym=j\Bm$ with $\Bm\in\mathbb{R}^{P\times P}$ and $\Bm=\Bm^\TT$.
\end{proposition}
\begin{proof}
See Appendix~\ref{app:class}.
\end{proof}

We call $\Bm$ the \emph{susceptance matrix} of the network, from which \eqref{eq:components} gives every tunable capacitor and inductor. A \ac{MiLAC} realizing \eqref{eq:hermplace} is neither lossless nor reciprocal. Indeed, its admittance matrix $\Ym=Y_0(\Pm-\Id)$ is Hermitian, so $\Ym+\Ym^\HH=2\Ym$, which is nonzero since the diagonal block $-Y_0(\Cm_x^{-1}+\Id)$ never vanishes, whereas a lossless network has $\Ym+\Ym^\HH=\Zerom$ (Appendix~\ref{app:class}). Unless $\Hm$, $\Cm_n$, and $\Cm_x$ are real, $\Ym$ is also not symmetric, so realizing \eqref{eq:hermplace} requires an unconstrained \ac{MiLAC}.

In contrast, a lossless and reciprocal \ac{MiLAC} realizes
\begin{equation}
\label{eq:Wlr}
\Wm=\Bigl[\Bigl(\frac{j\Bm}{Y_0}+\Id\Bigr)^{-1}\Bigr]_{21},\qquad \Bm=\Bm^\TT\in\mathbb{R}^{P\times P},
\end{equation}
with every component a tunable susceptance, the only resistive elements being the source and load admittances $Y_0$ \emph{outside} the network. Since such a network is passive, the gain that it can realize between disjoint input and output ports is bounded. To quantify this bound, we also describe the network by its \emph{scattering matrix} $\Sm\in\mathbb{C}^{P\times P}$. With $\mathbf a$ and $\mathbf b$ denoting the vectors of the waves incident on and reflected from the ports, referenced to $Y_0$, the port voltages and currents are \cite{Nerini2026Hybrid,Pozar2011Microwave}
\begin{equation}
\label{eq:waves}
\vv=\mathbf a+\mathbf b,\qquad \iv=Y_0(\mathbf a-\mathbf b),\qquad \mathbf b=\Sm\mathbf a,
\end{equation}
and $\Sm$ is unitary for a lossless network \cite{Nerini2025Capacity,Pozar2011Microwave}. We partition $\mathbf a$, $\mathbf b$, and $\Sm$ between the input and output ports as $\vv$ in \eqref{eq:interface}, and denote by $\Sm_{21}$ the output-by-input block of $\Sm$.

\begin{proposition}[Gain bound]
\label{prop:bound}
As in \cite{Nerini2025Capacity,Nerini2026Hybrid}, the output-by-input block \eqref{eq:Wblock} of a \ac{MiLAC} whose input and output ports are disjoint is
\begin{equation}
\label{eq:WS21}
\Wm=\frac12\Sm_{21}.
\end{equation}
If the \ac{MiLAC} is lossless, then $\|\Wm\|_2\le\tfrac12$.
\end{proposition}
\begin{proof}
See Appendix~\ref{app:class}.
\end{proof}

The factor $\tfrac12$ arises because each source launches half of its voltage into the network as an incident wave, and the unit bound on $\|\Sm_{21}\|_2$ expresses that a lossless network cannot amplify. Conversely, any matrix of spectral norm at most one is the block $\Sm_{21}$ of a unitary and symmetric scattering matrix \cite{Wu2026MiLAC}. The bound therefore constrains only the scale of a realized estimator, since any matrix satisfies it after multiplication by a small enough scalar. Accordingly, each realization below produces its target estimator multiplied by a known scaling factor, which is removed by a fixed gain in the \ac{RF} chains.

However, realizing a scaled estimator with the construction of \cite{Wu2026MiLAC} requires computing the estimator digitally first. The problem addressed in this paper can now be stated precisely: \emph{can the estimators \eqref{eq:rzf} and \eqref{eq:lmmse}, computed in prior work through \eqref{eq:hermplace} on unconstrained \acp{MiLAC}, be computed in the lossless and reciprocal class \eqref{eq:Wlr}, without digital matrix inversion when the channel changes?} Sections~\ref{sec:rzf} and \ref{sec:lmmse} answer it affirmatively by explicit construction for \ac{RLS} and \ac{LMMSE}, respectively.

\section{Lossless and Reciprocal MiLAC for RLS}
\label{sec:rzf}

In this section, we show how a lossless and reciprocal \ac{MiLAC} computes the \ac{RLS} estimator \eqref{eq:rzf}. We start with \ac{RLS} because the two weights of \eqref{eq:lmmse} reduce for \ac{RLS} to multiples of the identity, namely $\Cm_n^{-1}=\Id$ on the received signal and $\Cm_x^{-1}=\lambda\Id$, so the design involves only the channel. The main difficulty is that the channel is complex, whereas the susceptance matrix is real. The key idea is to double the ports, so that each complex signal is carried by a pair of ports and the channel enters the susceptance matrix through a real matrix of twice its size. We first describe the receiver in Section~\ref{sec:rzf-arch}. We then design the susceptance matrix and derive the output in Section~\ref{sec:rzf-net}, and finally give the algorithm and analyze its computational complexity in Section~\ref{sec:rzf-cost}.

\subsection{Receiver Architecture}
\label{sec:rzf-arch}

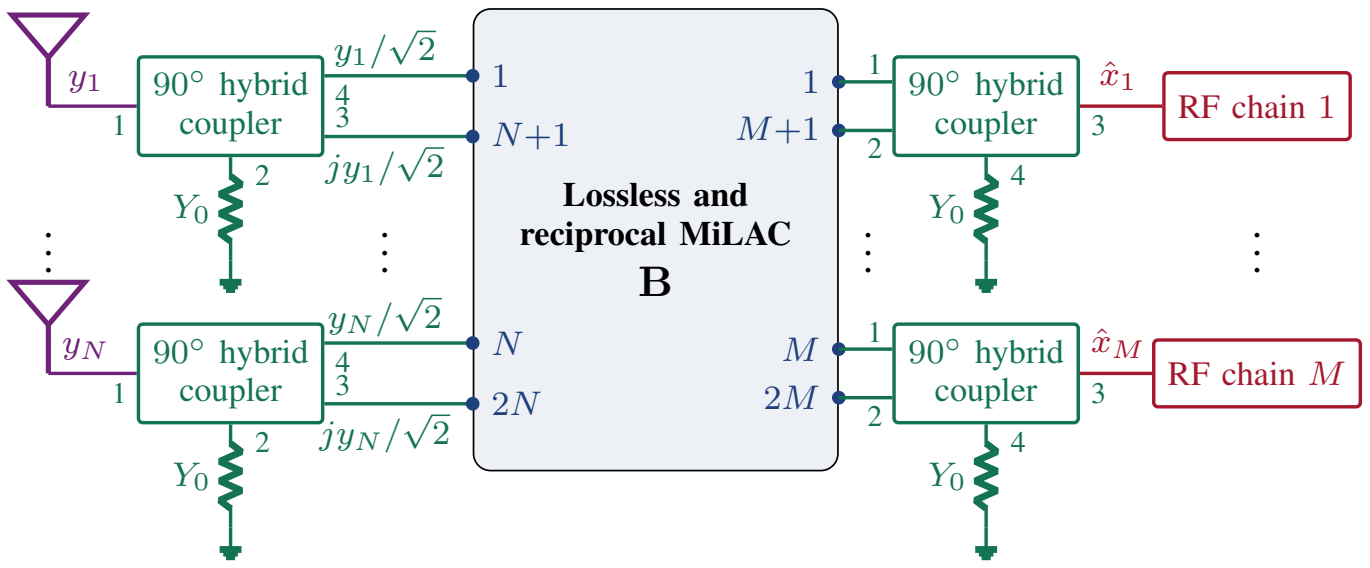
\begin{figure}[t]
\centering
\resizebox{\columnwidth}{!}{%
\ctikzset{bipoles/length=0.6cm}%
\begin{circuitikz}[transform shape, >=Latex, every node/.style={font=\footnotesize}]
\node[draw,rounded corners,fill=cport!8,minimum width=3.0cm,minimum height=3.8cm,align=center,font=\bfseries\footnotesize] (net) at (5.35,1.7){Lossless and\\ reciprocal MiLAC\\[3pt] \normalsize $\Bm$};
\foreach \y/\k/\i/\kk/\ii in {2.8/1/1/{N{+}1}/{M{+}1}, 0.6/N/M/{2N}/{2M}}{
  \draw[csrc,very thick] (0.35,\y)--(0.35,\y+0.4); \draw[csrc,very thick] (0.05,\y+0.75)--(0.35,\y+0.4)--(0.65,\y+0.75)--cycle;
  \node[draw=ccoup,thick,fill=white,rounded corners=1pt,minimum width=1.2cm,minimum height=0.72cm,align=center,font=\footnotesize,text=ccoup] (H\k) at (1.85,\y){$90^\circ$ hybrid\\ coupler};
  \draw[csrc,thick] (0.35,\y)--(H\k.west);
  \node[csrc,font=\footnotesize] at (0.66,\y+0.2){$y_{\k}$};
  \draw[ccoup,thick] (H\k.south) to[R,l_=$Y_0$] ($(H\k.south)+(0,-0.85)$) node[ground]{};
  \node[ccoup,font=\scriptsize,anchor=north east,inner sep=1pt] at ($(H\k.west)+(-0.04,-0.03)$){1}; \node[ccoup,font=\scriptsize,anchor=north west,inner sep=1pt] at ($(H\k.south)+(0.16,-0.02)$){2}; \node[ccoup,font=\scriptsize,anchor=north west,inner sep=1pt] at ($(H\k.east)+(0.05,0.22)$){4}; \node[ccoup,font=\scriptsize,anchor=south west,inner sep=1pt] at ($(H\k.east)+(0.05,-0.22)$){3};
  \coordinate (a\k) at (net.west|-0,\y+0.25); \coordinate (b\k) at (net.west|-0,\y-0.25); \coordinate (c\i) at (net.east|-0,\y+0.2); \coordinate (d\i) at (net.east|-0,\y-0.2);
  \draw[ccoup,thick] ($(H\k.east)+(0,0.25)$) -- (a\k); \draw[ccoup,thick] ($(H\k.east)+(0,-0.25)$) -- (b\k);
  \node[ccoup,font=\footnotesize] at (3.12,\y+0.47){$y_{\k}/\sqrt2$};
  \node[ccoup,font=\footnotesize] at (3.12,\y-0.49){$jy_{\k}/\sqrt2$};
  \filldraw[cport] (a\k) circle(1.4pt); \filldraw[cport] (b\k) circle(1.4pt);
  \node[cport,font=\footnotesize,anchor=west] at ($(a\k)+(0.03,0)$){$\k$};
  \node[cport,font=\footnotesize,anchor=west] at ($(b\k)+(0.03,0)$){$\kk$};
  \filldraw[cport] (c\i) circle(1.4pt); \filldraw[cport] (d\i) circle(1.4pt);
  \node[cport,font=\footnotesize,anchor=east] at ($(c\i)+(-0.03,0)$){$\i$};
  \node[cport,font=\footnotesize,anchor=east] at ($(d\i)+(-0.03,0)$){$\ii$};
  \node[draw=ccoup,thick,fill=white,rounded corners=1pt,minimum width=1.1cm,minimum height=0.72cm,align=center,font=\footnotesize,text=ccoup] (G\i) at (8.07,\y){$90^\circ$ hybrid\\ coupler};
  \draw[ccoup,thick] (c\i) -- (G\i.west|-c\i); \draw[ccoup,thick] (d\i) -- (G\i.west|-d\i);
  \node[ccoup,font=\scriptsize,anchor=south east,inner sep=1pt] at ($(G\i.west|-c\i)+(-0.04,0.02)$){1}; \node[ccoup,font=\scriptsize,anchor=north east,inner sep=1pt] at ($(G\i.west|-d\i)+(-0.04,-0.02)$){2};
  \node[ccoup,font=\scriptsize,anchor=north west,inner sep=1pt] at ($(G\i.east)+(0.05,-0.03)$){3}; \node[ccoup,font=\scriptsize,anchor=north west,inner sep=1pt] at ($(G\i.south)+(0.16,-0.02)$){4};
  \draw[ccoup,thick] (G\i.south) to[R,l_=$Y_0$] ($(G\i.south)+(0,-0.85)$) node[ground]{};
  \node[draw=cread,thick,fill=white,rounded corners=1pt,minimum width=1.2cm,minimum height=0.55cm,align=center,font=\footnotesize,text=cread] (R\i) at (10.3,\y){RF chain $\i$};
  \draw[cread,thick] (G\i.east) -- (R\i.west);
  \node[cread,font=\footnotesize] at (9.16,\y+0.25){$\hat x_{\i}$};
}
\node[font=\normalsize] at (0.35,1.7){$\vdots$}; \node[font=\normalsize] at (3.12,1.7){$\vdots$}; \node[font=\normalsize] at (7.1,1.7){$\vdots$}; \node[font=\normalsize] at (10.3,1.7){$\vdots$};
\end{circuitikz}}
\caption{Receiver for \ac{RLS}: The receiver comprises $N$ antennas, $N$ input hybrid couplers, a lossless and reciprocal \ac{MiLAC} with $2N$ input and $2M$ output ports, $M$ output hybrid couplers, and $M$ \ac{RF} chains. Small green numbers mark the ports of each hybrid coupler, and the numbers at the \ac{MiLAC} mark its input and output ports. The labels on the input lines are the equivalent source voltages of Lemma~\ref{lem:hybrid}, $\hat x_i$ marks the input of the \ac{RF} chain that receives the estimate of stream $i$ up to a known scaling factor, and all terminations are $Y_0$.}
\label{fig:archrzf}
\end{figure}

Consider the receiver in Fig.~\ref{fig:archrzf}, whose purpose is to deliver the estimate $\hat\xv=\Wm_{\mathrm{RLS}}\yv$ to the $M$ \ac{RF} chains, one stream per chain. With a pair of ports per complex signal, the \ac{MiLAC} has $2N$ input ports and $2M$ output ports. The design of Section~\ref{sec:rzf-net} uses, at the pair of input ports of each antenna, two copies of the antenna signal with a $90^\circ$ phase difference, and the input hybrid couplers provide these copies. Similarly, the output hybrid couplers combine each pair of output ports into one \ac{RF} chain. We index the input ports as $1,\dots,2N$ and the output ports as $1,\dots,2M$, so that output port $i$ is port $2N+i$ in the numbering of Fig.~\ref{fig:model}.

\emph{Antennas:} We denote by $y_k$ the phasor of the voltage induced at antenna $k$, i.e., the $k$th entry of $\yv=\Hm\xv+\nv$. As in \cite{Nerini2025AnalogII}, we assume that the antennas are perfectly matched to the reference impedance $Z_0=1/Y_0$, so that the Th\'evenin equivalent circuit of antenna $k$ is a voltage source of voltage $y_k$ in series with $Z_0$, as for the input ports in Fig.~\ref{fig:model}. As in most prior \ac{MiLAC} studies, we also neglect the mutual coupling between the antennas, whose effect is analyzed in \cite{Nerini2026Physics}.

\emph{Input hybrid couplers:} Antenna $k$ is connected to port $1$ of a $90^\circ$ hybrid coupler, a passive four-port component that divides the signal entering one port into two signals of equal amplitude and $90^\circ$ phase difference at two other ports \cite{Nerini2026Hybrid,Pozar2011Microwave}. Denote by $\alpha_p$ and $\beta_p$ the waves incident on and reflected from its port $p$, as in \eqref{eq:waves}. With the port numbering of \cite{Nerini2026Hybrid}, its scattering matrix is
\begin{equation}
\label{eq:hybridS}
\Sm_{\mathrm{hyb}}=\frac{1}{\sqrt2}\begin{bmatrix}0&0&j&1\\0&0&1&j\\ j&1&0&0\\ 1&j&0&0\end{bmatrix}.
\end{equation}
This matrix is unitary and symmetric, so the hybrid coupler is lossless and reciprocal \cite{Nerini2026Hybrid,Pozar2011Microwave}. Port $2$ of the hybrid coupler is terminated in $Y_0$, and ports $4$ and $3$ are connected to input ports $k$ and $N{+}k$ of the \ac{MiLAC}. The following lemma characterizes the resulting sources at these input ports.

\begin{lemma}[Sources produced by a hybrid coupler]
\label{lem:hybrid}
Let a voltage source of voltage $y_k$ in series with $Z_0$ be connected to port $1$ of the hybrid coupler \eqref{eq:hybridS}, and let port $2$ be terminated in $Y_0$. Then, regardless of the network connected to ports $4$ and $3$, these two ports are equivalent to two independent voltage sources $y_k/\sqrt2$ and $jy_k/\sqrt2$, each in series with $Z_0$.
\end{lemma}
\begin{proof}
See Appendix~\ref{app:hybrid}.
\end{proof}

Lemma~\ref{lem:hybrid} holds because ports $3$ and $4$ of the hybrid coupler are matched and isolated from each other. A wave reflected by the \ac{MiLAC} into port $3$ or $4$ therefore leaves through ports $1$ and $2$, where the matched antenna and the termination absorb it, and never returns to the \ac{MiLAC}. Input ports $k$ and $N{+}k$ of the \ac{MiLAC} are thus driven exactly as the input ports of Fig.~\ref{fig:model}, and the vector $\uv$ of source voltages in \eqref{eq:interface} is
\begin{equation}
\label{eq:quadsource}
\uv=\frac{1}{\sqrt2}\begin{bmatrix}\yv\\ j\yv\end{bmatrix},
\end{equation}
i.e., the first $N$ input ports receive the antenna signals scaled by $1/\sqrt2$, and the last $N$ input ports receive the same signals shifted in phase by $90^\circ$.

\emph{Output hybrid couplers and \ac{RF} chains:} Output ports $i$ and $M{+}i$ of the \ac{MiLAC} are connected to ports $1$ and $2$ of an output hybrid coupler \eqref{eq:hybridS}, whose port $3$ feeds \ac{RF} chain $i$ and whose port $4$ is terminated in $Y_0$. Since the input impedance of the \ac{RF} chain is matched to $Z_0$, ports $3$ and $4$ are matched, and by \eqref{eq:hybridS} ports $1$ and $2$ then reflect no wave. Each output port of the \ac{MiLAC} is therefore terminated in $Y_0$, as in Fig.~\ref{fig:model}. \ac{RF} chain $i$ amplifies, demodulates, and samples the signal that it receives. The two ports of a pair carry two carrier-frequency signals, not the in-phase and quadrature components of one signal. As in any receiver, the \ac{RF} chain alone forms these components.

\subsection{Design of the MiLAC and Its Output}
\label{sec:rzf-net}

We now design the susceptance matrix $\Bm$ of the \ac{MiLAC} in Fig.~\ref{fig:archrzf}. The \ac{RLS} estimator \eqref{eq:rzf} applies $\Hm^\HH$ to the received signal, so the design places $\Hm^\HH$ between the input and the output ports. However, the susceptances are real, whereas $\Hm^\HH$ is complex. We therefore let $\Hm^\HH$ enter through its \emph{real-valued representation}, which maps a complex matrix $\Am\in\mathbb{C}^{m\times n}$ to the real matrix
\begin{equation}
\label{eq:lift}
\lift{\Am}\triangleq\begin{bmatrix}\realp{\Am}&\imagp{\Am}\\-\imagp{\Am}&\realp{\Am}\end{bmatrix}\in\mathbb{R}^{2m\times2n}.
\end{equation}
Up to a permutation of rows and columns, $\lift{\Am}$ replaces each complex entry of $\Am$ by a $2\times2$ real block containing its real and imaginary parts. The properties of $\lift{\cdot}$, labeled \ref{P:lin}--\ref{P:norm}, are collected in Appendix~\ref{app:lift}. With this representation, we set the susceptance matrix to
\begin{equation}
\label{eq:Brzf}
\Bm=Y_0\begin{bmatrix}\Zerom_{2N}&\Fm^\TT\\ \Fm&\Zerom_{2M}\end{bmatrix},\qquad \Fm=\frac{1}{\sqrt\lambda}\,\lift{\Hm^\HH},
\end{equation}
with $\Fm\in\mathbb{R}^{2M\times2N}$, which we call the \emph{channel block}. The channel block and its transpose are the only nonzero blocks of $\Bm$, so the network is bipartite: every inter-port susceptance joins an input port to an output port. By \eqref{eq:components}, the inter-port susceptance between input port $k$ and output port $i$ is $-Y_0[\Fm]_{i,k}$, and the port-to-ground susceptance of each port is the sum of the entries in its column of $\Bm$.

As a result, each complex entry $h$ of $\Hm^\HH$ is realized by four susceptances between a pair of input ports and a pair of output ports. Two properties of the representation explain why these real susceptances compute the complex estimate. First, by \ref{P:pair}, the block of $h$ maps the pair $(z,jz)$ that an input hybrid coupler delivers to the pair $(hz,jhz)$, whose first entry is the complex product. Second, since $\lift{\Am^\HH}=\lift{\Am}^\TT$ by \ref{P:herm}, a reciprocal network, whose susceptance matrix is symmetric, couples the output ports back to the input ports through $\Fm^\TT=\lift{\Hm}/\sqrt\lambda$. The network thus contains both $\Hm^\HH$ and $\Hm$, the two factors of the Gram matrix $\Hm^\HH\Hm$ in \eqref{eq:rzf}. Although no regularization term appears in \eqref{eq:Brzf}, the following theorem shows that the terminations of the output ports provide the term $\lambda\Id$ of \eqref{eq:rzf} and that these ports carry the \ac{RLS} estimate.

\begin{theorem}[Output of the RLS receiver]
\label{thm:rzf}
Let $\Hm\in\mathbb{C}^{N\times M}$, $\lambda>0$, and $\Wm_{\mathrm{RLS}}=(\Hm^\HH\Hm+\lambda\Id_M)^{-1}\Hm^\HH$. The lossless and reciprocal \ac{MiLAC} with susceptance matrix \eqref{eq:Brzf}, its input ports driven through $Y_0$ and its output ports terminated in $Y_0$ as in Fig.~\ref{fig:model}, realizes the output-by-input block
\begin{equation}
\label{eq:rzfmap}
\bigl[\Pm^{-1}\bigr]_{21}=-j\sqrt\lambda\;\lift{\Wm_{\mathrm{RLS}}},
\end{equation}
and, in the receiver of Fig.~\ref{fig:archrzf}, driven by \eqref{eq:quadsource}, the voltages at its output ports are
\begin{equation}
\label{eq:rzfout}
\vv_2=-j\sqrt{\frac{\lambda}{2}}\begin{bmatrix}\Wm_{\mathrm{RLS}}\yv\\ j\Wm_{\mathrm{RLS}}\yv\end{bmatrix}.
\end{equation}
The realized block \eqref{eq:rzfmap} satisfies the bound of Proposition~\ref{prop:bound}, since $2\sqrt\lambda\,\|\Wm_{\mathrm{RLS}}\|_2\le1$, with equality if and only if $\lambda$ is the square of a singular value of $\Hm$.
\end{theorem}
\begin{proof}
See Appendix~\ref{app:rzf}.
\end{proof}

Theorem~\ref{thm:rzf} shows that output port $i$ carries the phasor $-j\sqrt{\lambda/2}\,[\Wm_{\mathrm{RLS}}\yv]_i$, i.e., the estimate of stream $i$ multiplied by a known scaling factor, and that output port $M{+}i$ carries the same estimate shifted in phase by $90^\circ$. Output hybrid coupler $i$ combines these two copies. Since its ports $1$ and $2$ reflect no wave, \eqref{eq:waves} gives the waves incident on these ports as $\alpha_1=[\vv_2]_i$ and $\alpha_2=[\vv_2]_{M+i}=j[\vv_2]_i$. By \eqref{eq:hybridS}, its ports $4$ and $3$ then reflect $\beta_4=(\alpha_1+j\alpha_2)/\sqrt2=0$ and $\beta_3=(j\alpha_1+\alpha_2)/\sqrt2=\sqrt2\,j[\vv_2]_i$. The termination at port $4$ thus receives no power, and \ac{RF} chain $i$ receives all the output signal power. Since the chain is matched, its input voltage is $\beta_3=\sqrt\lambda\,[\Wm_{\mathrm{RLS}}\yv]_i$, and a fixed gain $1/\sqrt\lambda$ in the chain removes this factor. The receiver of Fig.~\ref{fig:archrzf} thus delivers $\hat\xv=\Wm_{\mathrm{RLS}}\yv$ to its $M$ \ac{RF} chains without computing it digitally. Since the \ac{MiLAC} and the hybrid couplers are lossless and reciprocal, the only resistive elements are the antenna source impedances, the terminations, and the \ac{RF} chain inputs.

To gain insight into Theorem~\ref{thm:rzf}, consider the structure of $\Pm=j\Bm/Y_0+\Id$. Its diagonal blocks are the identities contributed by the admittances $Y_0$, and its off-diagonal blocks, $j\Fm$ and, by reciprocity, $j\Fm^\TT$, couple the input and output ports only through the channel block. Block matrix inversion, carried out in Appendix~\ref{app:rzf}, gives $[\Pm^{-1}]_{21}=-(\Id+\Fm\Fm^\TT)^{-1}(j\Fm)$, where $\Id+\Fm\Fm^\TT=\lift{\lambda\Id+\Hm^\HH\Hm}/\lambda$. Hence, the identity contributed by the terminations of the output ports provides the regularization $\lambda\Id$, and the network inverts $\lambda\Id+\Hm^\HH\Hm$ as the signals propagate.

Theorem~\ref{thm:rzf} also covers the two limiting cases of \ac{RLS} identified in Section~\ref{sec:place}, as stated in the following corollary.

\begin{corollary}[ZF and MF]
\label{cor:limits}
As $\lambda\to0$, the estimator $\Wm_{\mathrm{RLS}}$ realized by the design \eqref{eq:Brzf} tends to the \ac{ZF} estimator $(\Hm^\HH\Hm)^{-1}\Hm^\HH$ if $\Hm$ has full column rank. As $\lambda\to\infty$, $\lambda\Wm_{\mathrm{RLS}}$ tends to the \ac{MF} $\Hm^\HH$.
\end{corollary}
\begin{proof}
See Appendix~\ref{app:rzf}.
\end{proof}

However, neither limit can be reached in hardware. As $\lambda\to\infty$, the inter-port susceptances vanish, since $\|\Fm\|_2=\|\Hm\|_2/\sqrt\lambda$ by \ref{P:norm}. The output ports then decouple from the input ports, so the output voltages \eqref{eq:rzfout} vanish, and recovering the \ac{MF} output $\Hm^\HH\yv$ requires an unbounded gain in the \ac{RF} chains. As $\lambda\to0$, the inter-port susceptances instead grow without bound, which tunable components of finite range cannot realize.

\subsection{Algorithm and Computational Complexity}
\label{sec:rzf-cost}

Algorithm~\ref{alg:rzf} summarizes the design of Section~\ref{sec:rzf-net}. We now evaluate the number of real operations that a digital processor performs to configure and operate the \ac{MiLAC}, and compare it with that of a digital \ac{RLS} receiver. Two timescales are involved: the channel $\Hm$ is fixed within each coherence block, and each coherence block contains many symbols.

\emph{Per coherence block:} The only computation is to rewrite the channel block $\Fm=\lift{\Hm^\HH}/\sqrt\lambda$, whose entries are the real and imaginary parts of the entries of $\Hm$ scaled by $\pm1/\sqrt\lambda$, and to convert it into component values by \eqref{eq:components}. This costs $O(MN)$ real operations and involves no matrix product or inversion, since the network forms and inverts $\Hm^\HH\Hm+\lambda\Id$ as the signals propagate. A digital receiver must instead form $\Hm^\HH\Hm+\lambda\Id$ at $O(M^2N)$ and invert it at $O(M^3)$ in every coherence block.

\emph{Per symbol:} The \ac{MiLAC} requires no digital arithmetic, whereas a digital receiver computes $\hat\xv=\Wm_{\mathrm{RLS}}\yv$ at $O(MN)$.

Unlike \ac{LMMSE} in Section~\ref{sec:lmmse}, \ac{RLS} requires no offline computation, since its weights $\Id$ and $\lambda\Id$ need no factorization. Moreover, the unconstrained \ac{MiLAC} computing \ac{RLS} has the same per-block and per-symbol costs, since it also rewrites only the channel-dependent entries of its admittance matrix, at $O(MN)$ \cite{Nerini2025AnalogII}. The lossless and reciprocal constraint therefore adds no computation to leading order.

\begin{algorithm}[t]
\caption{Design of the \ac{MiLAC} for \ac{RLS}}
\label{alg:rzf}
\begin{algorithmic}[1]
\Require Channel $\Hm$, regularization weight $\lambda>0$, and $Y_0$.
\Statex \textit{Per coherence block:}
\State Compute the channel block $\Fm\gets\lift{\Hm^\HH}/\sqrt\lambda$.
\State Assemble $\Bm$ by \eqref{eq:Brzf} and set the susceptances by \eqref{eq:components}.
\Statex \textit{Per symbol:}
\State Apply $\uv$ in \eqref{eq:quadsource} through the input hybrid couplers.
\State Receive $\sqrt\lambda\,\hat x_i$ at \ac{RF} chain $i$ through output hybrid coupler $i$ and apply the gain $1/\sqrt\lambda$, for $i=1,\dots,M$.
\Ensure Estimate $\hat\xv=\Wm_{\mathrm{RLS}}\yv$ at the \ac{RF} chains.
\end{algorithmic}
\end{algorithm}

\section{Lossless and Reciprocal MiLAC for LMMSE}
\label{sec:lmmse}

In this section, we extend the receiver of Section~\ref{sec:rzf} to the \ac{LMMSE} estimator \eqref{eq:lmmse}. The \ac{LMMSE} estimator differs from \ac{RLS} in two weights: the received signal is weighted by the inverse noise covariance $\Cm_n^{-1}$ instead of $\Id$, and the regularization $\lambda\Id$ is replaced by the inverse prior covariance $\Cm_x^{-1}$. In Section~\ref{sec:lmmse-arch}, we show that these weights require conductances, which susceptances cannot provide, and we obtain them from auxiliary ports. Sections~\ref{sec:lmmse-net} and \ref{sec:lmmse-cost} then have the same structure as Sections~\ref{sec:rzf-net} and \ref{sec:rzf-cost}.

\subsection{Receiver Architecture}
\label{sec:lmmse-arch}

We first explain why the \ac{RLS} design cannot produce the \ac{LMMSE} weights $\Cm_n^{-1}$ and $\Cm_x^{-1}$. In that design, the diagonal blocks of $\Pm$ are the identities contributed by the admittances $Y_0$. The input identity gives the weight $\Id$, and the output identity, combined with the factor $1/\sqrt\lambda$ in $\Fm$, gives the regularization $\lambda\Id$. For \ac{LMMSE}, the analogous design needs an input block proportional to $\lift{\Cm_n}$, whose inverse gives the weight $\Cm_n^{-1}$, and an output block proportional to $\lift{\Cm_x^{-1}}$. Both blocks are real and, in general, differ from the identity. Without additional ports, however, susceptances cannot produce them, since $j\Bm$ is purely imaginary and the real part of $\Pm=j\Bm/Y_0+\Id$ is thus the identity contributed by the admittances $Y_0$. These blocks therefore require additional conductances, which a lossless network does not contain.

\begin{figure}[t]
\centering
\resizebox{\columnwidth}{!}{%
\ctikzset{bipoles/length=0.6cm}%
\begin{circuitikz}[transform shape, >=Latex, every node/.style={font=\footnotesize}]
\node[draw,rounded corners,fill=cport!8,minimum width=3.0cm,minimum height=3.8cm,align=center,font=\bfseries\footnotesize] (net) at (5.35,1.7){Lossless and\\ reciprocal MiLAC\\[3pt] \normalsize $\Bm$};
\foreach \x/\cc in {4.0/cshunt,4.35/cshunt,4.85/cshunt,5.85/caux,6.2/caux,6.7/caux}{
  \draw[\cc,thick] (\x,3.6) -- (\x,3.72) to[R] (\x,4.42) node[ground,rotate=180]{};
  \filldraw[\cc] (\x,3.6) circle(1.4pt);}
\node[cshunt,font=\normalsize] at (4.6,4.07){$\cdots$};
\node[caux,font=\normalsize] at (6.45,4.07){$\cdots$};
\draw[cshunt,thick,decorate,decoration={brace,amplitude=4pt}] (3.88,4.72) -- (4.97,4.72);
\draw[caux,thick,decorate,decoration={brace,amplitude=4pt}] (5.73,4.72) -- (6.82,4.72);
\node[cshunt,align=center] at (4.42,5.2){Noise stage\\ $m_{\mathrm D}$ ports};
\node[caux,align=center] at (6.27,5.2){Prior stage\\ $m_{\mathrm E}$ ports};
\foreach \y/\k/\i/\kk/\ii in {2.8/1/1/{N{+}1}/{M{+}1}, 0.6/N/M/{2N}/{2M}}{
  \draw[csrc,very thick] (0.35,\y)--(0.35,\y+0.4); \draw[csrc,very thick] (0.05,\y+0.75)--(0.35,\y+0.4)--(0.65,\y+0.75)--cycle;
  \node[draw=ccoup,thick,fill=white,rounded corners=1pt,minimum width=1.2cm,minimum height=0.72cm,align=center,font=\footnotesize,text=ccoup] (H\k) at (1.85,\y){$90^\circ$ hybrid\\ coupler};
  \draw[csrc,thick] (0.35,\y)--(H\k.west);
  \node[csrc,font=\footnotesize] at (0.66,\y+0.2){$y_{\k}$};
  \draw[ccoup,thick] (H\k.south) to[R,l_=$Y_0$] ($(H\k.south)+(0,-0.85)$) node[ground]{};
  \node[ccoup,font=\scriptsize,anchor=north east,inner sep=1pt] at ($(H\k.west)+(-0.04,-0.03)$){1}; \node[ccoup,font=\scriptsize,anchor=north west,inner sep=1pt] at ($(H\k.south)+(0.16,-0.02)$){2}; \node[ccoup,font=\scriptsize,anchor=north west,inner sep=1pt] at ($(H\k.east)+(0.05,0.22)$){4}; \node[ccoup,font=\scriptsize,anchor=south west,inner sep=1pt] at ($(H\k.east)+(0.05,-0.22)$){3};
  \coordinate (a\k) at (net.west|-0,\y+0.25); \coordinate (b\k) at (net.west|-0,\y-0.25); \coordinate (c\i) at (net.east|-0,\y+0.2); \coordinate (d\i) at (net.east|-0,\y-0.2);
  \draw[ccoup,thick] ($(H\k.east)+(0,0.25)$) -- (a\k); \draw[ccoup,thick] ($(H\k.east)+(0,-0.25)$) -- (b\k);
  \node[ccoup,font=\footnotesize] at (3.12,\y+0.47){$y_{\k}/\sqrt2$};
  \node[ccoup,font=\footnotesize] at (3.12,\y-0.49){$jy_{\k}/\sqrt2$};
  \filldraw[cport] (a\k) circle(1.4pt); \filldraw[cport] (b\k) circle(1.4pt);
  \node[cport,font=\footnotesize,anchor=west] at ($(a\k)+(0.03,0)$){$\k$};
  \node[cport,font=\footnotesize,anchor=west] at ($(b\k)+(0.03,0)$){$\kk$};
  \filldraw[cport] (c\i) circle(1.4pt); \filldraw[cport] (d\i) circle(1.4pt);
  \node[cport,font=\footnotesize,anchor=east] at ($(c\i)+(-0.03,0)$){$\i$};
  \node[cport,font=\footnotesize,anchor=east] at ($(d\i)+(-0.03,0)$){$\ii$};
  \node[draw=ccoup,thick,fill=white,rounded corners=1pt,minimum width=1.1cm,minimum height=0.72cm,align=center,font=\footnotesize,text=ccoup] (G\i) at (8.07,\y){$90^\circ$ hybrid\\ coupler};
  \draw[ccoup,thick] (c\i) -- (G\i.west|-c\i); \draw[ccoup,thick] (d\i) -- (G\i.west|-d\i);
  \node[ccoup,font=\scriptsize,anchor=south east,inner sep=1pt] at ($(G\i.west|-c\i)+(-0.04,0.02)$){1}; \node[ccoup,font=\scriptsize,anchor=north east,inner sep=1pt] at ($(G\i.west|-d\i)+(-0.04,-0.02)$){2};
  \node[ccoup,font=\scriptsize,anchor=north west,inner sep=1pt] at ($(G\i.east)+(0.05,-0.03)$){3}; \node[ccoup,font=\scriptsize,anchor=north west,inner sep=1pt] at ($(G\i.south)+(0.16,-0.02)$){4};
  \draw[ccoup,thick] (G\i.south) to[R,l_=$Y_0$] ($(G\i.south)+(0,-0.85)$) node[ground]{};
  \node[draw=cread,thick,fill=white,rounded corners=1pt,minimum width=1.2cm,minimum height=0.55cm,align=center,font=\footnotesize,text=cread] (R\i) at (10.3,\y){RF chain $\i$};
  \draw[cread,thick] (G\i.east) -- (R\i.west);
  \node[cread,font=\footnotesize] at (9.16,\y+0.25){$\hat x_{\i}$};
}
\node[font=\normalsize] at (0.35,1.7){$\vdots$}; \node[font=\normalsize] at (3.12,1.7){$\vdots$}; \node[font=\normalsize] at (7.1,1.7){$\vdots$}; \node[font=\normalsize] at (10.3,1.7){$\vdots$};
\end{circuitikz}}
\caption{Receiver for \ac{LMMSE}: The receiver of Fig.~\ref{fig:archrzf} with a noise stage of $m_{\mathrm D}$ and a prior stage of $m_{\mathrm E}$ auxiliary ports, each terminated in $Y_0$.}
\label{fig:archlmmse}
\end{figure}
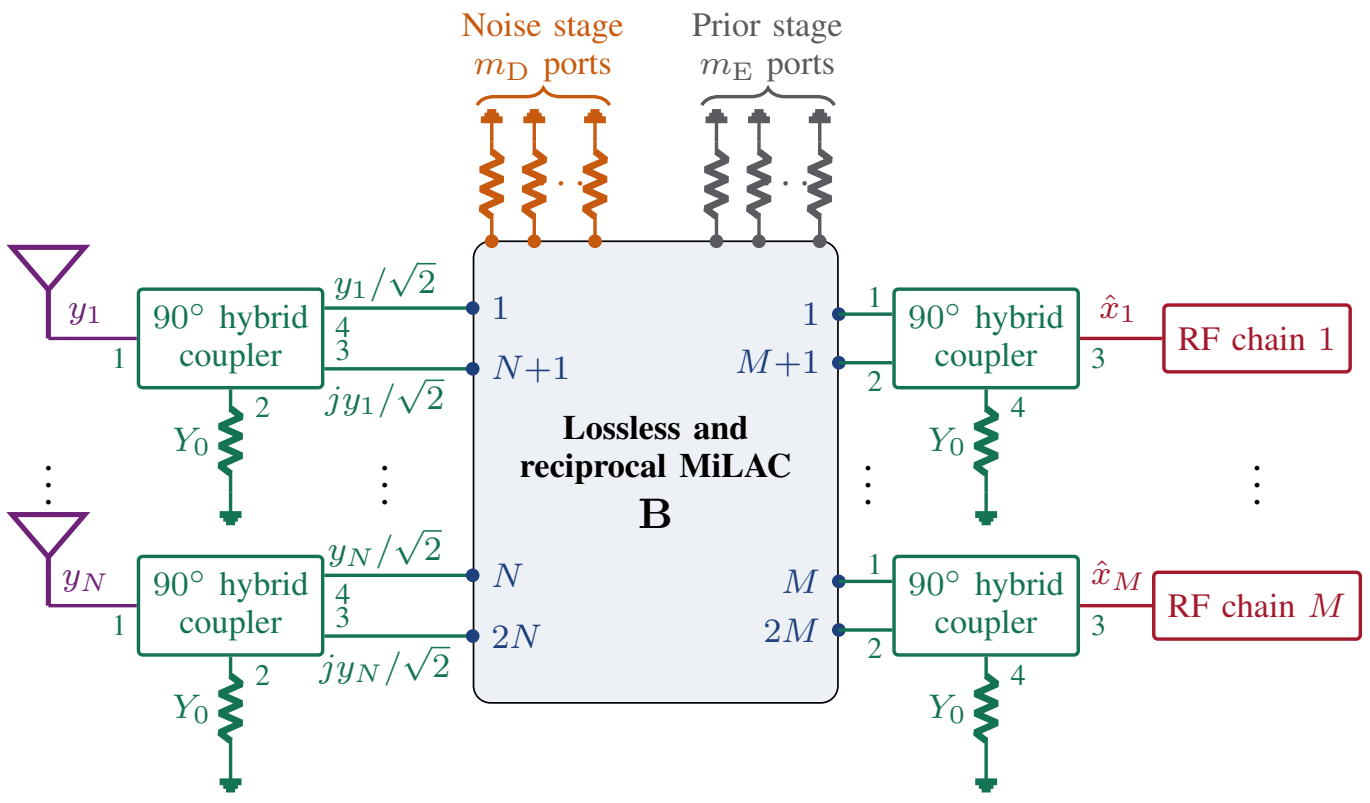

To obtain these conductances, the receiver in Fig.~\ref{fig:archlmmse} equips its \ac{MiLAC} with additional ports. Besides the input and output ports of Fig.~\ref{fig:archrzf}, the \ac{MiLAC} has $m_{\mathrm D}+m_{\mathrm E}$ \emph{auxiliary} ports. An auxiliary port is coupled to input or output ports by tunable susceptances, and it is terminated in $Y_0$. However, unlike the input and output ports, it is neither driven by a source nor connected to an \ac{RF} chain. The $m_{\mathrm D}$ auxiliary ports coupled to the input ports, together with their coupling susceptances, form the \emph{noise stage}, which produces the weight $\Cm_n^{-1}$. Likewise, the $m_{\mathrm E}$ auxiliary ports coupled to the output ports form the \emph{prior stage}, which produces the weight $\Cm_x^{-1}$.

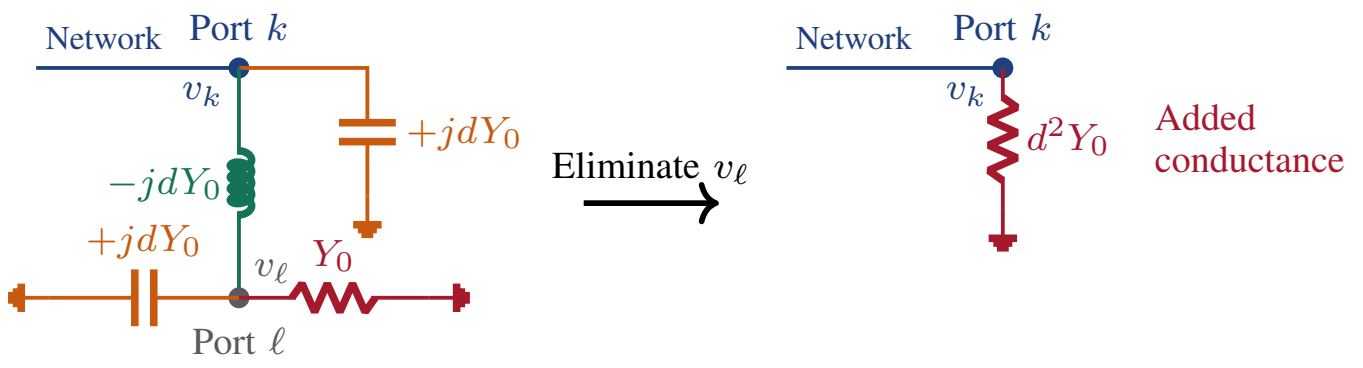
\begin{figure}[t]
\centering
\resizebox{\columnwidth}{!}{%
\ctikzset{bipoles/length=0.7cm}%
\begin{circuitikz}[transform shape, every node/.style={font=\footnotesize}]
\draw[cport,thick] (-1.5,0) -- (0,0) coordinate(k);
\node[cport,font=\scriptsize] at (-1.0,0.22){Network};
\filldraw[cport] (k) circle(2pt); \node[cport] at (0,0.3){Port $k$}; \node[cport,anchor=north east,inner sep=1pt] at (-0.07,-0.07){$v_k$};
\draw[cshunt,thick] (k) -- (0.95,0) to[C,l=$+jdY_0$] (0.95,-0.95) node[ground]{};
\draw[ccoup,thick] (k) to[L,l_=$-jdY_0$] (0,-1.7) coordinate(a);
\filldraw[caux] (a) circle(2pt); \node[caux] at (0,-2.02){Port $\ell$}; \node[caux,anchor=south west,inner sep=1pt] at (0.07,-1.63){$v_\ell$};
\draw[cshunt,thick] (a) to[C,l_=$+jdY_0$] (-1.4,-1.7) node[rotate=-90,ground]{};
\draw[cread,thick] (a) to[R,l=$Y_0$] (1.4,-1.7) node[rotate=90,ground]{};
\draw[->,very thick] (2.55,-1.0)--(3.55,-1.0) node[midway,above,font=\footnotesize]{Eliminate $v_\ell$};
\begin{scope}[shift={(5.35,0)}]
\draw[cport,thick] (-1.3,0) -- (0.3,0) coordinate(k2);
\node[cport,font=\scriptsize] at (-0.8,0.22){Network};
\filldraw[cport] (k2) circle(2pt); \node[cport] at (0.3,0.3){Port $k$}; \node[cport,anchor=north east,inner sep=1pt] at (0.23,-0.07){$v_k$};
\draw[cread,thick] (k2) to[R,l=$d^2Y_0$] (0.3,-1.05) node[ground]{};
\node[cread,anchor=west,align=left] at (1.3,-0.52){Added\\ conductance};
\end{scope}
\end{circuitikz}}
\caption{Toy example of a terminated auxiliary port. Left: Auxiliary port $\ell$, terminated in $Y_0$, is coupled to port $k$ by $[\Bm]_{k,\ell}=[\Bm]_{\ell,k}=dY_0$. Right: Equivalent circuit at port $k$ after the voltage $v_\ell$ is eliminated. Component labels are admittances, and for $d>0$ the inter-port component is an inductor and the port-to-ground components are capacitors.}
\label{fig:reduction}
\end{figure}

\emph{A toy example:} To see how the termination of an auxiliary port adds conductance at the ports that it is coupled to, although the \ac{MiLAC} contains only susceptances, consider the simplest case in Fig.~\ref{fig:reduction}. A single auxiliary port $\ell$ is coupled to a single port $k$ through $[\Bm]_{k,\ell}=[\Bm]_{\ell,k}=dY_0$, for a real number $d$, with $[\Bm]_{k,k}=[\Bm]_{\ell,\ell}=0$. By \eqref{eq:components}, this coupling is realized by an inter-port component of admittance $-jdY_0$ between the two ports, and it contributes $+jdY_0$ to the port-to-ground component of each of them. In the row of port $\ell$, the matrix $\Pm=j\Bm/Y_0+\Id$ has the entries $[\Pm]_{\ell,k}=jd$, from the coupling, and $[\Pm]_{\ell,\ell}=1$, from the termination $Y_0$. Since port $\ell$ has no source, the row of $\Pm\vv=\tilde\uv$ for port $\ell$ reads
\begin{equation}
\label{eq:auxrow}
jd\,v_k+v_\ell=0 .
\end{equation}
The voltage of port $\ell$, marked in Fig.~\ref{fig:reduction}, is thus $v_\ell=-jd\,v_k$. Substituting it into the row of port $k$ eliminates $v_\ell$ and turns the coupling term into
\begin{equation}
\label{eq:realincr}
[\Pm]_{k,\ell}v_\ell=(jd)(-jd)\,v_k=d^2v_k.
\end{equation}
This term is proportional to $v_k$, so after the elimination, the row of port $k$ is that of a network without port $\ell$ in which $[\Pm]_{k,k}$ is increased by $d^2$. Since $\Pm=\Ym/Y_0+\Id$, this increase is the same as connecting a conductance $d^2Y_0$ from port $k$ to ground, as in Fig.~\ref{fig:reduction}, right. The added admittance is real, although every component is a susceptance, for two reasons. The coupling enters twice, through $[\Pm]_{\ell,k}$ and $[\Pm]_{k,\ell}$, and $j^2=-1$. The termination of port $\ell$ gives the real entry $[\Pm]_{\ell,\ell}=1$.

\subsection{Design of the MiLAC and Its Output}
\label{sec:lmmse-net}

We now design the susceptance matrix $\Bm$ of the \ac{MiLAC} in Fig.~\ref{fig:archlmmse}, auxiliary ports included, with the ports ordered as input, noise-stage, output, and prior-stage ports. We choose the proportionality constants of Section~\ref{sec:lmmse-arch} as scalars $s\ge1/\lmin(\Cm_n)$ and $\eta\ge\lmax(\Cm_x)$, and set $g=\sqrt{s\eta}$, as justified after \eqref{eq:Peff}. We then take the channel block $\Fm$ and real factors $\Dm\in\mathbb{R}^{2N\times m_{\mathrm D}}$ and $\Em\in\mathbb{R}^{2M\times m_{\mathrm E}}$ satisfying
\begin{align}
\Dm\Dm^\TT&=s\lift{\Cm_n}-\Id,\label{eq:DEFlmmse}\\
\Fm&=g\lift{\Hm^\HH},\label{eq:Flmmse}\\
\Em\Em^\TT&=\eta\lift{\Cm_x^{-1}}-\Id.\label{eq:Elmmse}
\end{align}
By \ref{P:lin} and \ref{P:spec}, the right-hand sides of \eqref{eq:DEFlmmse} and \eqref{eq:Elmmse} equal $\lift{s\Cm_n-\Id}$ and $\lift{\eta\Cm_x^{-1}-\Id}$, which are positive semidefinite if and only if $s\lmin(\Cm_n)\ge1$ and $\eta\lmin(\Cm_x^{-1})=\eta/\lmax(\Cm_x)\ge1$. The choice of $s$ and $\eta$ ensures both conditions, so such factors exist. With these factors, we set the susceptance matrix to
\begin{equation}
\label{eq:Btemplate}
\Bm=Y_0\begin{bmatrix}\Zerom & \Dm & \Fm^\TT & \Zerom\\ \Dm^\TT & \Zerom & \Zerom & \Zerom\\ \Fm & \Zerom & \Zerom & \Em\\ \Zerom & \Zerom & \Em^\TT & \Zerom\end{bmatrix}.
\end{equation}
In \eqref{eq:Btemplate}, the channel block $\Fm$ couples the input ports to the output ports as in \eqref{eq:Brzf}, the factor $\Dm$ couples the input ports to the noise-stage ports, and the factor $\Em$ couples the output ports to the prior-stage ports. The auxiliary ports are coupled to nothing else. The factor $\Dm$ depends on $\Cm_n$ alone, $\Em$ on $\Cm_x$ alone, and \emph{only $\Fm$ depends on the channel $\Hm$}, so only $\Fm$ is rewritten when the channel changes, as Section~\ref{sec:lmmse-cost} quantifies.

The numbers $m_{\mathrm D}$ and $m_{\mathrm E}$ of auxiliary ports are the numbers of columns of the factors. They can be taken equal to the ranks of the right-hand sides of \eqref{eq:DEFlmmse} and \eqref{eq:Elmmse}, which by \ref{P:norm} are twice those of $s\Cm_n-\Id$ and $\eta\Cm_x^{-1}-\Id$, and hence at most $2N$ and $2M$. The smallest admissible values, $s=1/\lmin(\Cm_n)$ and $\eta=\lmax(\Cm_x)$, make both matrices singular and thus reduce the number of auxiliary ports, and Algorithm~\ref{alg:lmmse} uses them. As shown in Section~\ref{sec:hw}, $m_{\mathrm D}$ is then much smaller than $2N$ for white noise plus low-rank interference.

The elimination of the toy example also applies to the groups of auxiliary ports in \eqref{eq:Btemplate}. The noise-stage ports are coupled to the input ports through the real matrix $\Dm$ in place of $d$, and eliminating their voltages adds $\Dm\Dm^\TT$ to the identity block of the input ports, which becomes $s\lift{\Cm_n}$ by \eqref{eq:DEFlmmse}. Likewise, the prior stage adds $\Em\Em^\TT$ to the identity block of the output ports, which becomes $\eta\lift{\Cm_x^{-1}}$ by \eqref{eq:Elmmse}. The input and output ports are therefore described exactly by the reduced matrix
\begin{equation}
\label{eq:Peff}
\Pm_{\mathrm{eff}}=\begin{bmatrix}s\lift{\Cm_n} & j\Fm^\TT\\ j\Fm & \eta\lift{\Cm_x^{-1}}\end{bmatrix},
\end{equation}
as shown in Step~1 of the proof in Appendix~\ref{app:gen}. This matrix has the form of the matrix $\Pm$ of the \ac{RLS} design, with the identity blocks replaced by scaled representations of $\Cm_n$ and $\Cm_x^{-1}$. The reduced matrix also explains the conditions on $s$ and $\eta$. Indeed, the admittances $Y_0$ at the input and output ports contribute the identity to both diagonal blocks of \eqref{eq:Peff}, and a terminated auxiliary port can only add conductance, as $d^2Y_0\ge0$ in Fig.~\ref{fig:reduction}. The stages thus add the positive semidefinite matrices $\Dm\Dm^\TT$ and $\Em\Em^\TT$ to these identities, so that neither diagonal block can have an eigenvalue smaller than $1$, which by \ref{P:spec} requires $s\ge1/\lmin(\Cm_n)$ and $\eta\ge\lmax(\Cm_x)$.

The output of the design then follows from block matrix inversion of \eqref{eq:Peff}, carried out in Appendix~\ref{app:gen}. The inverse of the input block $s\lift{\Cm_n}$ weights the received signal by $\Cm_n^{-1}$. The Schur complement of this block contains $(g^2/s)\lift{\Hm^\HH\Cm_n^{-1}\Hm}$, the scaled representation of the weighted Gram matrix. The choice $g=\sqrt{s\eta}$ makes this scale equal to that of the prior block $\eta\lift{\Cm_x^{-1}}$, so the two terms add to $\eta\lift{\Hm^\HH\Cm_n^{-1}\Hm+\Cm_x^{-1}}$, the representation of the matrix that the \ac{LMMSE} estimator inverts. Neither this matrix nor its inverse is computed digitally. The following theorem gives the resulting output.

\begin{theorem}[Output of the LMMSE receiver]
\label{thm:lmmse}
Let $\Hm\in\mathbb{C}^{N\times M}$, $\Cm_n\succ\Zerom$, $\Cm_x\succ\Zerom$, and $\Wm_{\mathrm{LMMSE}}=(\Hm^\HH\Cm_n^{-1}\Hm+\Cm_x^{-1})^{-1}\Hm^\HH\Cm_n^{-1}$, and denote by $[\Pm^{-1}]_{\mathrm{oi}}$ the submatrix of $\Pm^{-1}$ with rows at the output ports and columns at the input ports, the counterpart of $[\Pm^{-1}]_{21}$ in Theorem~\ref{thm:rzf}. With $s\ge1/\lmin(\Cm_n)$, $\eta\ge\lmax(\Cm_x)$, $g=\sqrt{s\eta}$, and $\Dm$, $\Fm$, and $\Em$ as in \eqref{eq:DEFlmmse}--\eqref{eq:Elmmse}, the lossless and reciprocal \ac{MiLAC} with susceptance matrix \eqref{eq:Btemplate}, its input ports driven through $Y_0$ and all other ports terminated in $Y_0$ as in Fig.~\ref{fig:archlmmse}, realizes
\begin{equation}
\label{eq:lmmsemap}
\bigl[\Pm^{-1}\bigr]_{\mathrm{oi}}=-\frac{j}{\sqrt{s\eta}}\,\lift{\Wm_{\mathrm{LMMSE}}},
\end{equation}
and, in the receiver of Fig.~\ref{fig:archlmmse}, driven by \eqref{eq:quadsource}, the voltages at its output ports are
\begin{equation}
\label{eq:lmmseout}
\vv_{\mathrm o}=-\frac{j}{\sqrt{2s\eta}}\begin{bmatrix}\Wm_{\mathrm{LMMSE}}\yv\\ j\Wm_{\mathrm{LMMSE}}\yv\end{bmatrix}.
\end{equation}
\end{theorem}
\begin{proof}
See Appendix~\ref{app:gen}.
\end{proof}

Theorem~\ref{thm:lmmse} shows that output port $i$ carries the estimate $[\Wm_{\mathrm{LMMSE}}\yv]_i$ multiplied by the known scaling factor $-j/\sqrt{2s\eta}$, and that output port $M{+}i$ carries the same estimate shifted in phase by $90^\circ$. As for \ac{RLS}, \ac{RF} chain $i$ thus receives $[\Wm_{\mathrm{LMMSE}}\yv]_i/\sqrt{s\eta}$, and a fixed gain $\sqrt{s\eta}$ removes this factor.

\ac{RLS} is the special case of white noise and an isotropic prior. With $\Cm_n=\Id$ and $\Cm_x=\lambda^{-1}\Id$, the values $s=1$ and $\eta=1/\lambda$ give $g=1/\sqrt\lambda$ and empty factors, $m_{\mathrm D}=m_{\mathrm E}=0$, so \eqref{eq:Btemplate} reduces to \eqref{eq:Brzf} and the scaling factor in \eqref{eq:lmmsemap} to that of Theorem~\ref{thm:rzf}. Compared with the \ac{RLS} receiver, the only added resistive elements are the terminations of the auxiliary ports, and the \ac{MiLAC} itself remains lossless and reciprocal.

\begin{algorithm}[t]
\caption{Design of the \ac{MiLAC} for \ac{LMMSE}}
\label{alg:lmmse}
\begin{algorithmic}[1]
\Require Channel $\Hm$, covariances $\Cm_n,\Cm_x\succ\Zerom$, and $Y_0$.
\Statex \textit{Offline, once per change of $\Cm_n$ and $\Cm_x$:}
\State Set $s\gets 1/\lmin(\Cm_n)$, $\eta\gets\lmax(\Cm_x)$, and $g\gets\sqrt{s\eta}$.
\State Compute $\Dm$ and $\Em$ satisfying \eqref{eq:DEFlmmse} and \eqref{eq:Elmmse}, with as many columns as the ranks of their right-hand sides.
\Statex \textit{Per coherence block:}
\State Compute the channel block $\Fm\gets g\lift{\Hm^\HH}$.
\State Assemble $\Bm$ by \eqref{eq:Btemplate} and set the susceptances by \eqref{eq:components}.
\Statex \textit{Per symbol:}
\State Apply $\uv$ in \eqref{eq:quadsource} through the input hybrid couplers.
\State Receive $\hat x_i/\sqrt{s\eta}$ at \ac{RF} chain $i$ through output hybrid coupler $i$ and apply the gain $\sqrt{s\eta}$, for $i=1,\dots,M$.
\Ensure Estimate $\hat\xv=\Wm_{\mathrm{LMMSE}}\yv$ at the \ac{RF} chains.
\end{algorithmic}
\end{algorithm}

\subsection{Algorithm and Computational Complexity}
\label{sec:lmmse-cost}

Algorithm~\ref{alg:lmmse} summarizes the design of Section~\ref{sec:lmmse-net}. We evaluate its number of real operations as in Section~\ref{sec:rzf-cost}, now over three nested timescales: the covariances $\Cm_n$ and $\Cm_x$ are fixed over many coherence blocks, and the channel $\Hm$ is fixed within each coherence block, which contains many symbols.

\emph{Offline, once per change of the covariances:} The factors $\Dm$ and $\Em$ and the inverse $\Cm_x^{-1}$ cost $O(M^3+N^3)$. By \eqref{eq:DEFlmmse} and \eqref{eq:Elmmse}, they are independent of $\Hm$.

\emph{Per coherence block:} Only the channel block $\Fm=g\lift{\Hm^\HH}$ is rewritten and converted into component values by \eqref{eq:components}, at $O(MN)$ as for \ac{RLS}, and no matrix product or inversion is computed. A digital receiver must instead form $\Wm_{\mathrm{LMMSE}}$ in each coherence block, at $O(M^3+M^2N+MN^2)$ even with $\Cm_n^{-1}$ and $\Cm_x^{-1}$ precomputed offline. This number is dominated by the inverse-covariance weighting $\Cm_n^{-1}\Hm$ when $N\gg M$.

\emph{Per symbol:} As for \ac{RLS}, the \ac{MiLAC} requires no digital arithmetic, whereas a digital receiver computes $\hat\xv=\Wm_{\mathrm{LMMSE}}\yv$ at $O(MN)$.

Table~\ref{tab:complexity} summarizes the comparison for both estimators and also includes the unconstrained \ac{MiLAC} of \cite{Nerini2025AnalogI}, whose matrix $\Pm$ in \eqref{eq:hermplace} contains $\Cm_n$ directly. Offline, this \ac{MiLAC} therefore needs only $\Cm_x^{-1}$ and the $O(N^2)$ components of the $\Cm_n$ block, and per coherence block, it rewrites its channel-dependent block $\Hm$ at the same $O(MN)$ \cite{Nerini2025AnalogI}. Hence, to leading order, the lossless and reciprocal \ac{MiLAC} achieves the same saving over the digital receiver, per coherence block and per symbol, as the unconstrained \ac{MiLAC}.

\begin{remark}[Why the advantage does not extend to matrix inversion]
\label{rem:inv}
With the constructions of this paper, the per-block advantage of Table~\ref{tab:complexity} is specific to estimation and does not extend to general matrix inversion. An unconstrained \ac{MiLAC} inverts an $L\times L$ matrix by writing it directly into its admittance matrix, at $O(L^2)$ operations against the $O(L^3)$ of a digital inversion \cite{Nerini2025AnalogI}. In contrast, a lossless and reciprocal \ac{MiLAC} cannot directly write a matrix whose real part differs from the identity, since the real part of $\Pm$ is the identity contributed by the admittances $Y_0$. Coupling to other ports, each loaded by $Y_0$, can add to this identity only the product of a real matrix with its transpose, such as $\Fm\Fm^\TT$ in Section~\ref{sec:rzf-net} or $\Dm\Dm^\TT$ in Section~\ref{sec:lmmse-net}. Obtaining such a factor of a general matrix, for example by a Cholesky decomposition, is itself an $O(L^3)$ computation, so for a single inversion, the cubic cost is shifted to the factorization rather than removed. The estimators avoid this cost because only the covariances are factored, once and offline. The channel, the only quantity that changes in each coherence block, enters the matrix to be inverted only through $\Hm^\HH$ and $\Hm$, and $\Hm^\HH$ is written directly into $\Fm$ at $O(MN)$.
\end{remark}

\section{Circuit Complexity Analysis}
\label{sec:hw}

In addition to the computational complexity, the practicality of a \ac{MiLAC} depends on its \emph{circuit complexity}, defined as its number of tunable components \cite{Nerini2026Stem}. In this section, we derive the circuit complexity of each realization, which is determined by the sparsity pattern of its susceptance matrix $\Bm$.

By \eqref{eq:components}, each nonzero entry of $\Bm$ above the diagonal is one inter-port susceptance, and each port has one port-to-ground susceptance. If the network is modeled as a graph whose vertices are the ports and whose edges are the inter-port susceptances, a \ac{MiLAC} with $N_V$ ports and $N_E$ edges uses \cite{Nerini2026Stem}
\begin{equation}
\label{eq:Ncount}
N_C=\underbrace{N_V}_{\text{Port-to-ground}}+\underbrace{N_E}_{\text{Inter-port}}
\end{equation}
tunable components. In other words, the sparsity pattern of $\Bm$ is the topology of the circuit: ports $m$ and $n$ are coupled if and only if $[\Bm]_{m,n}\ne0$. A \emph{fully-connected} \ac{MiLAC} couples every pair of ports, i.e., $N_E=N_V(N_V-1)/2$, so by \eqref{eq:Ncount} its number of components \cite{Nerini2026Stem}
\begin{equation}
\label{eq:Nfull}
N_C^{\mathrm{full}}=\frac{N_V(N_V+1)}{2}=O(N_V^2)
\end{equation}
grows quadratically with the number of ports. For the large arrays considered here, such a growth is impractical, which has motivated sparser topologies whose number of components grows linearly \cite{Nerini2026Stem}.

We now show that, for the \ac{RLS} realization \eqref{eq:Brzf} and the \ac{LMMSE} realization \eqref{eq:Btemplate}, the channel block yields a bipartite topology whose number of components is linear in $N$. Only the noise stage can have a number of components that grows quadratically in $N$, and only for an unstructured noise covariance rather than white noise plus low-rank interference. The resulting numbers of components are collected in Table~\ref{tab:hardware}.

\begin{table}[t]
\centering
\caption{Real operations at each timescale of a digital receiver, the unconstrained \ac{MiLAC} of \cite{Nerini2025AnalogI,Nerini2025AnalogII}, and the lossless and reciprocal \ac{MiLAC}, for $M$ streams and $N$ antennas.}
\label{tab:complexity}
\renewcommand{\arraystretch}{1.25}
\setlength{\tabcolsep}{3.5pt}
\footnotesize
\resizebox{\columnwidth}{!}{%
\begin{tabular}{@{}llccc@{}}
\toprule
Timescale & Estimator & Digital & \begin{tabular}[c]{@{}c@{}}Unconstrained\\ \ac{MiLAC} \cite{Nerini2025AnalogI,Nerini2025AnalogII}\end{tabular} & This work \\
\midrule
Offline & \ac{RLS} & None & None & None \\
 & \ac{LMMSE} & $O(M^3{+}N^3)$ & $O(M^3{+}N^2)$ & $O(M^3{+}N^3)$ \\
Per block & \ac{RLS} & $O(M^3{+}M^2N)$ & $O(MN)$ & $O(MN)$ \\
 & \ac{LMMSE} & $O(M^3{+}M^2N{+}MN^2)$ & $O(MN)$ & $O(MN)$ \\
Per symbol & Both & $O(MN)$ & None & None \\
\bottomrule
\end{tabular}}
\end{table}

\subsection{Topology of the Channel Block}
\label{sec:hw-bip}

The following proposition gives the graph of the \ac{RLS} realization \eqref{eq:Brzf} and its number of components.

\begin{proposition}[\ac{RLS} topology and number of components]
\label{prop:rzf-hw}
For a generic $\Hm$, for which every entry of $\Fm$ and every column sum of $\Bm$ is nonzero, the \ac{RLS} realization \eqref{eq:Brzf} has a complete bipartite graph, no auxiliary ports, $N_E=4MN$ inter-port susceptances, and $N_V=2M+2N$ port-to-ground susceptances, which by \eqref{eq:Ncount} give the number of components
\begin{equation}
\label{eq:Nrzf}
N_C^{\mathrm{RLS}}=4MN+2M+2N=O(MN),
\end{equation}
linear in $N$ for a fixed $M$, against $O((M+N)^2)$ for a fully-connected \ac{MiLAC} on the same ports.
\end{proposition}
\begin{proof}
See Appendix~\ref{app:hw}.
\end{proof}

The same block $\Fm$ contains the channel in the \ac{LMMSE} matrix \eqref{eq:Btemplate}, where, by \eqref{eq:DEFlmmse}--\eqref{eq:Elmmse}, it is the only channel-dependent block. The reconfiguration performed in each coherence block thus affects only these $O(MN)$ components. We next analyze the blocks $\Dm$ and $\Em$ of the noise and prior stages.

\begin{figure*}[t]
\centering
\resizebox{\textwidth}{!}{%
\begin{tikzpicture}[font=\footnotesize,>=Latex,
  inp/.style={circle,draw=cport!70!black,fill=cport!18,minimum size=2.9mm,inner sep=0pt},
  nsp/.style={circle,draw=cshunt!70!black,fill=cshunt!30,minimum size=2.9mm,inner sep=0pt}]
\def\cs{0.25}
\def\ct{1.95}     
\def\cb{-2.30}    
\def\yhh{1.84}    
\def\yg{0.78}     
\def\ym{-0.12}    
\def\yc{-2.02}    
\def\xM{-4.45}\def\xa{-0.6}\def\xb{3.0}\def\xd{6.6}
\def\hw{1.66}
\fill[gray!11,rounded corners=2pt] (\xM-1.5,\cb) rectangle (\xM+1.45,\ct);
\begin{scope}[shift={(\xM-0.82,0.55)}]
\def\mb{0.5}
\fill[cshunt!42] (1*\mb,0) rectangle ++(\mb,-\mb);
\fill[cshunt!42] (0,-1*\mb) rectangle ++(\mb,-\mb);
\fill[ccoup!38] (2*\mb,0) rectangle ++(\mb,-\mb);
\fill[ccoup!38] (0,-2*\mb) rectangle ++(\mb,-\mb);
\fill[caux!40] (3*\mb,-2*\mb) rectangle ++(\mb,-\mb);
\fill[caux!40] (2*\mb,-3*\mb) rectangle ++(\mb,-\mb);
\draw[gray!55,step=\mb] (0,0) grid (4*\mb,-4*\mb);
\node[font=\scriptsize,cshunt!80!black] at (1.5*\mb,-0.5*\mb){$\Dm$};
\node[font=\scriptsize,cshunt!80!black] at (0.5*\mb,-1.5*\mb){$\Dm^{\TT}$};
\node[font=\scriptsize,ccoup!60!black] at (2.5*\mb,-0.5*\mb){$\Fm^{\TT}$};
\node[font=\scriptsize,ccoup!60!black] at (0.5*\mb,-2.5*\mb){$\Fm$};
\node[font=\scriptsize,caux!80!black] at (3.5*\mb,-2.5*\mb){$\Em$};
\node[font=\scriptsize,caux!80!black] at (2.5*\mb,-3.5*\mb){$\Em^{\TT}$};
\foreach \k/\lab in {0/{$2N$},1/{$m_{\mathrm D}$},2/{$2M$},3/{$m_{\mathrm E}$}}{
  \node[font=\scriptsize,gray!40!black] at ({(\k+0.5)*\mb},0.2){\lab};
  \node[font=\scriptsize,gray!40!black,anchor=east] at (-0.03,{-(\k+0.5)*\mb}){\lab};}
\draw[cshunt!85!black,very thick,rounded corners=1pt] (0.96*\mb,0.04) rectangle (2.04*\mb,-0.96*\mb);
\end{scope}
\node[font=\small\bfseries,anchor=north] at (\xM,\yhh){$\Bm/Y_0$};
\node[font=\scriptsize,cshunt!70!black,align=center] at (\xM,\cb+0.36){Only $\Dm$ can grow\\ quadratically in $N$};
\node[font=\large,gray!55!black] at (\xM+1.85,-0.2){$\Rightarrow$};
\foreach \x in {\xa,\xb,\xd}{\fill[gray!6,rounded corners=2pt] (\x-\hw,\cb) rectangle (\x+\hw,\ct);}
\node[font=\small\bfseries,anchor=north] at (\xa,\yhh){White};
\node[font=\small\bfseries,anchor=north] at (\xb,\yhh){Rank-$r$ interference};
\node[font=\small\bfseries,anchor=north] at (\xd,\yhh){Unstructured};
\begin{scope}[shift={(\xa,\yg)}]
  \foreach \k in {0,...,4}{\node[inp] at ({(\k-2)*0.46},-0.3){};}
  \node[font=\scriptsize,gray!40!black] at (0,0.3){No auxiliary ports};
\end{scope}
\begin{scope}[shift={(\xb,\yg)}]
  \foreach \k in {0,...,4}{\node[inp] (r\k) at ({(\k-2)*0.46},-0.3){};}
  \foreach \j/\xx in {1/-0.23,2/0.23}{\node[nsp] (h\j) at (\xx,0.34){};}
  \foreach \k in {0,...,4}{\foreach \j in {1,2}{\draw[cshunt!70!black,line width=0.35pt](h\j)--(r\k);}}
\end{scope}
\begin{scope}[shift={(\xd,\yg)}]
  \foreach \k in {0,...,4}{\node[inp] (d\k) at ({(\k-2)*0.46},-0.3){};}
  \foreach \j in {0,...,3}{\node[nsp] (e\j) at ({(\j-1.5)*0.46},0.34){};}
  \foreach \k in {0,...,4}{\foreach \j in {0,...,3}{\draw[cshunt!70!black,line width=0.25pt](e\j)--(d\k);}}
\end{scope}
\begin{scope}[shift={(\xa-2.5*\cs,\ym)}]
  \draw[gray!50,step=\cs](0,0) grid (5*\cs,-5*\cs);
  \node[font=\scriptsize] at (2.5*\cs,0.17){$s\Cm_n-\Id=\Zerom$};
\end{scope}
\begin{scope}[shift={(\xb-1.12,\ym)}]
  \fill[gray!30] (0,0) rectangle (5*\cs,-5*\cs); \draw[gray!55,step=\cs](0,0) grid (5*\cs,-5*\cs);
  \node[font=\scriptsize] at (2.5*\cs,0.17){$s\Cm_n-\Id$};
  \node[font=\scriptsize] at (2.5*\cs,-5*\cs-0.2){Rank $r$};
  \node[font=\normalsize] at (5*\cs+0.25,-2.5*\cs){$\to$};
  \begin{scope}[shift={(5*\cs+0.5,0)}]
    \fill[cshunt!50] (0,0) rectangle (2*\cs,-5*\cs); \draw[gray!55,step=\cs](0,0) grid (2*\cs,-5*\cs);
    \node[font=\scriptsize,cshunt!80!black] at (1*\cs,0.17){$\Dm_{\mathrm c}$};
  \end{scope}
\end{scope}
\begin{scope}[shift={(\xd-1.38,\ym)}]
  \fill[gray!30] (0,0) rectangle (5*\cs,-5*\cs); \draw[gray!55,step=\cs](0,0) grid (5*\cs,-5*\cs);
  \node[font=\scriptsize] at (2.5*\cs,0.17){$s\Cm_n-\Id$};
  \node[font=\scriptsize] at (2.5*\cs,-5*\cs-0.2){Rank $\le N{-}1$};
  \node[font=\normalsize] at (5*\cs+0.25,-2.5*\cs){$\to$};
  \begin{scope}[shift={(5*\cs+0.5,0)}]
    \fill[cshunt!50] (0,0) rectangle (4*\cs,-5*\cs); \draw[gray!55,step=\cs](0,0) grid (4*\cs,-5*\cs);
    \node[font=\scriptsize,cshunt!80!black] at (2*\cs,0.17){$\Dm_{\mathrm c}$};
  \end{scope}
\end{scope}
\node[font=\small] at (\xa,\yc){No components};
\node[font=\small] at (\xb,\yc){$O(Nr)$ components};
\node[font=\small] at (\xd,\yc){$O(N^2)$ components};
\end{tikzpicture}}
\caption{Left: Block pattern of the \ac{LMMSE} susceptance matrix $\Bm/Y_0$ in \eqref{eq:Btemplate}, with rows and columns ordered as the $2N$ input, $m_{\mathrm D}$ noise-stage, $2M$ output, and $m_{\mathrm E}$ prior-stage ports. Right: The noise-stage block $\Dm$ for white noise, white noise plus rank-$r$ interference, and an unstructured noise covariance, with $N=5$ antennas and $r=2$. In each right panel, the gray matrix is $s\Cm_n-\Id$ and the orange matrix is its complex factor $\Dm_{\mathrm c}$, with $\Dm=\lift{\Dm_{\mathrm c}}$. Each blue node is an antenna, that is, a row of $\Dm_{\mathrm c}$, each orange node is a column of $\Dm_{\mathrm c}$, and each edge is a nonzero entry of $\Dm_{\mathrm c}$. The bottom line of each right panel gives the number of components of the noise stage.}
\label{fig:Bstruct}
\end{figure*}
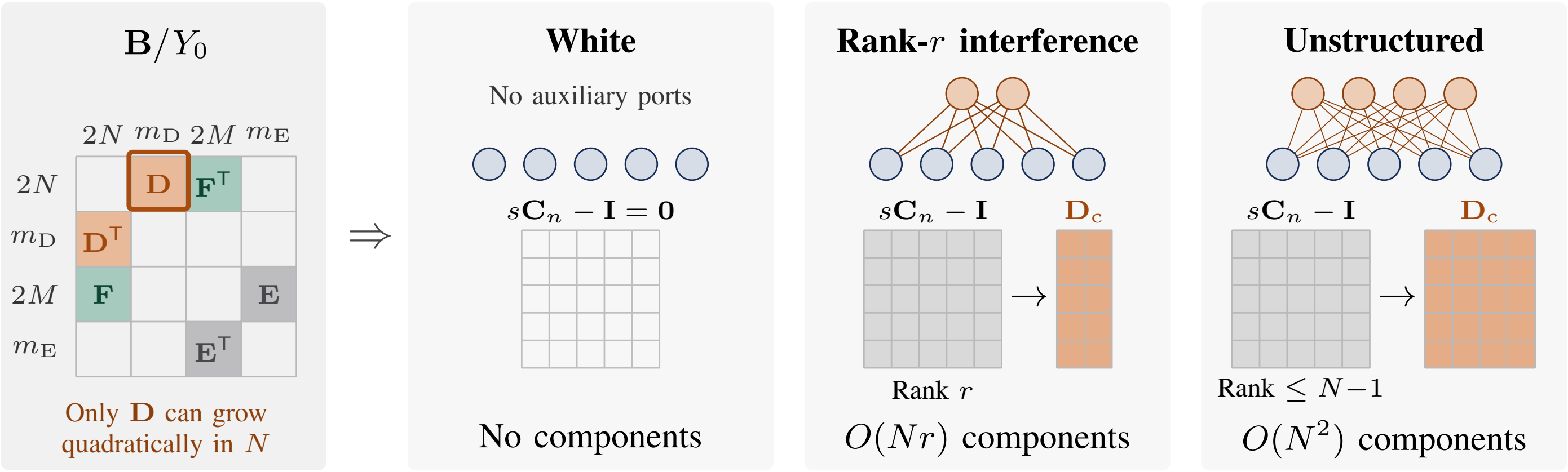

\subsection{Topology of the Prior and Noise Stages}
\label{sec:hw-lmmse}

Since the design fixes only the products $\Dm\Dm^\TT$ and $\Em\Em^\TT$ in \eqref{eq:DEFlmmse} and \eqref{eq:Elmmse}, not the factors themselves, the two stages need few components when their factors have few columns and few nonzero entries. With the minimal $\eta$ of Algorithm~\ref{alg:lmmse}, $\eta\Cm_x^{-1}-\Id$ is singular and has rank at most $M-1$, so it equals $\Em_{\mathrm c}\Em_{\mathrm c}^\HH$ for a complex factor $\Em_{\mathrm c}$ with at most $M-1$ columns. The factor $\Em=\lift{\Em_{\mathrm c}}$ satisfies \eqref{eq:Elmmse} by \ref{P:lin}, \ref{P:prod}, and \ref{P:herm}, and it gives two prior-stage ports per column of $\Em_{\mathrm c}$, hence at most $2M-2$. Moreover, multiplying a column of $\Em_{\mathrm c}$ by a unit-modulus number does not change $\Em_{\mathrm c}\Em_{\mathrm c}^\HH$, so we make the first entry of each column real. This removes one nonzero entry from each of the two columns of $\Em$ that the column generates. Each prior-stage port is thus coupled to at most $2M-1$ of the $2M$ output ports and has one port-to-ground component, so the prior stage has at most $4M(M-1)$ components, which is negligible for $M\ll N$. For the noise stage, the following proposition obtains such a factor from a factor of the complex matrix $s\Cm_n-\Id$.

\begin{proposition}[Number of noise-stage components]
\label{prop:lmmse-hw}
Any real factor $\Dm\in\mathbb{R}^{2N\times m_{\mathrm D}}$ satisfying \eqref{eq:DEFlmmse} realizes the noise stage with $m_{\mathrm D}$ auxiliary ports and one inter-port susceptance per nonzero entry of $\Dm$. For $M\le N$, taking $\Dm=\lift{\Dm_{\mathrm c}}$, with $\Dm_{\mathrm c}\Dm_{\mathrm c}^\HH=s\Cm_n-\Id$, gives the following three cases.
\begin{enumerate}[label=(\roman*),leftmargin=2.2em]
\item White noise, $\Cm_n=\sigma^2\Id$: With $s=1/\sigma^2$, $s\Cm_n-\Id=\Zerom$, so no noise stage is needed, and the number of components is $O(MN)$, as for \ac{RLS}.
\item White noise plus rank-$r$ interference, $\Cm_n=\sigma^2\Id+\Um\Um^\HH$ with $\Um\in\mathbb{C}^{N\times r}$: The value $s=1/\sigma^2$ is admissible since $\lmin(\Cm_n)\ge\sigma^2$, and $s\Cm_n-\Id=\sigma^{-2}\Um\Um^\HH$ has rank at most $r$. A $2N\times2r$ factor gives $2r$ auxiliary ports, and the number of components is $O(MN+Nr)$.
\item Unstructured $\Cm_n$: With $s=1/\lmin(\Cm_n)$, $s\Cm_n-\Id$ has rank at most $N-1$. A factor with as many columns as this rank gives at most $2N-2$ auxiliary ports, and the number of components is $O(N^2)$.
\end{enumerate}
\end{proposition}
\begin{proof}
See Appendix~\ref{app:hw}.
\end{proof}

Fig.~\ref{fig:Bstruct} illustrates Proposition~\ref{prop:lmmse-hw}. With the factor of Proposition~\ref{prop:lmmse-hw}, $\Dm$ has twice as many columns, and thus noise-stage ports, as the rank of $s\Cm_n-\Id$. Two observations follow. First, in case (ii), the value $s=1/\sigma^2$ minimizes the number of noise-stage ports. For any larger admissible $s$, the matrix $s\Cm_n-\Id$ has full rank, and its factors, obtained by a Cholesky or eigenvalue decomposition, give a $2N\times2N$ block $\Dm$ with $O(N^2)$ nonzero entries. Second, the saving in case (ii) involves no approximation of $\Cm_n$. Although the covariance $\Cm_n=\sigma^2\Id+\Um\Um^\HH$ is itself dense, the factor $\Dm=\sigma^{-1}\lift{\Um}$ satisfies \eqref{eq:DEFlmmse} exactly with only $2r$ auxiliary ports. Hence, only an unstructured noise covariance makes the number of components grow quadratically (Table~\ref{tab:hardware}), whereas white noise plus interference from a few sources keeps it linear.

\begin{table}[t]
\centering
\caption{Number of tunable components of the realizations, with $s$ as in Proposition~\ref{prop:lmmse-hw} and the minimal $\eta$ of Algorithm~\ref{alg:lmmse}, and of a fully-connected \ac{MiLAC} on the same $2M+2N$ ports, for $M$ streams, $N\ge M$ antennas, and interference rank $r$.}
\label{tab:hardware}
\renewcommand{\arraystretch}{1.3}
\setlength{\tabcolsep}{4pt}
\resizebox{\columnwidth}{!}{%
\begin{tabular}{@{}llcc@{}}
\toprule
Realization & Noise-stage ports & Components & Growth in $N$ \\
\midrule
Fully-connected \ac{MiLAC} & -- & $O((M{+}N)^2)$ & Quadratic \\
\ac{RLS} & None & $O(MN)$ & Linear \\
\ac{LMMSE}, white $\Cm_n$ & None & $O(MN)$ & Linear \\
\ac{LMMSE}, rank-$r$ interference & $2r$ & $O(MN{+}Nr)$ & Linear \\
\ac{LMMSE}, unstructured $\Cm_n$ & At most $2N{-}2$ & $O(N^2)$ & Quadratic \\
\bottomrule
\end{tabular}}
\end{table}

\section{Numerical Results}
\label{sec:sim}

In this section, we present numerical results to evaluate the cost of the proposed \acp{MiLAC}. By Theorems~\ref{thm:rzf} and \ref{thm:lmmse}, the \acp{MiLAC} of Sections~\ref{sec:rzf} and \ref{sec:lmmse} compute the \ac{RLS} and \ac{LMMSE} estimators exactly, so they attain the same performance as the corresponding digital receivers. Section~\ref{sec:sim-comp} evaluates the \emph{computational complexity}, i.e., the number of real operations performed digitally, against a digital receiver. Section~\ref{sec:sim-circ} evaluates the circuit complexity against a fully-connected \ac{MiLAC}, and Section~\ref{sec:sim-rate} reports the sum rate against the computational complexity.

\subsection{Computational Complexity}
\label{sec:sim-comp}

We evaluate the number of real operations per coherence block, i.e., over the $\tau$ symbols during which the channel is fixed. A complex addition costs two real operations and a complex multiplication six \cite{Nerini2025AnalogI}, so a complex multiply--accumulate costs eight. Per coherence block, a digital \ac{LMMSE} receiver computes the inverse-covariance weighting $\Cm_n^{-1}\Hm$ at $8MN^2$ for an unstructured $\Cm_n$, forms $\Km$ at $8M^2N$, inverts it at $\tfrac83M^3$ \cite{Nerini2025AnalogI}, and applies the estimator to the $\tau$ symbols at $8\tau MN$. The final formation of $\Wm_{\mathrm{LMMSE}}$ and the addition of $\Cm_x^{-1}$ to $\Km$ are not charged, so this number understates the digital cost by $O(M^2N)$. Evaluated in the same way, the digital \ac{RLS} receiver costs $8M^2N+\tfrac83M^3+8\tau MN$, and the digital \ac{MF} receiver costs $8\tau MN$, per coherence block.

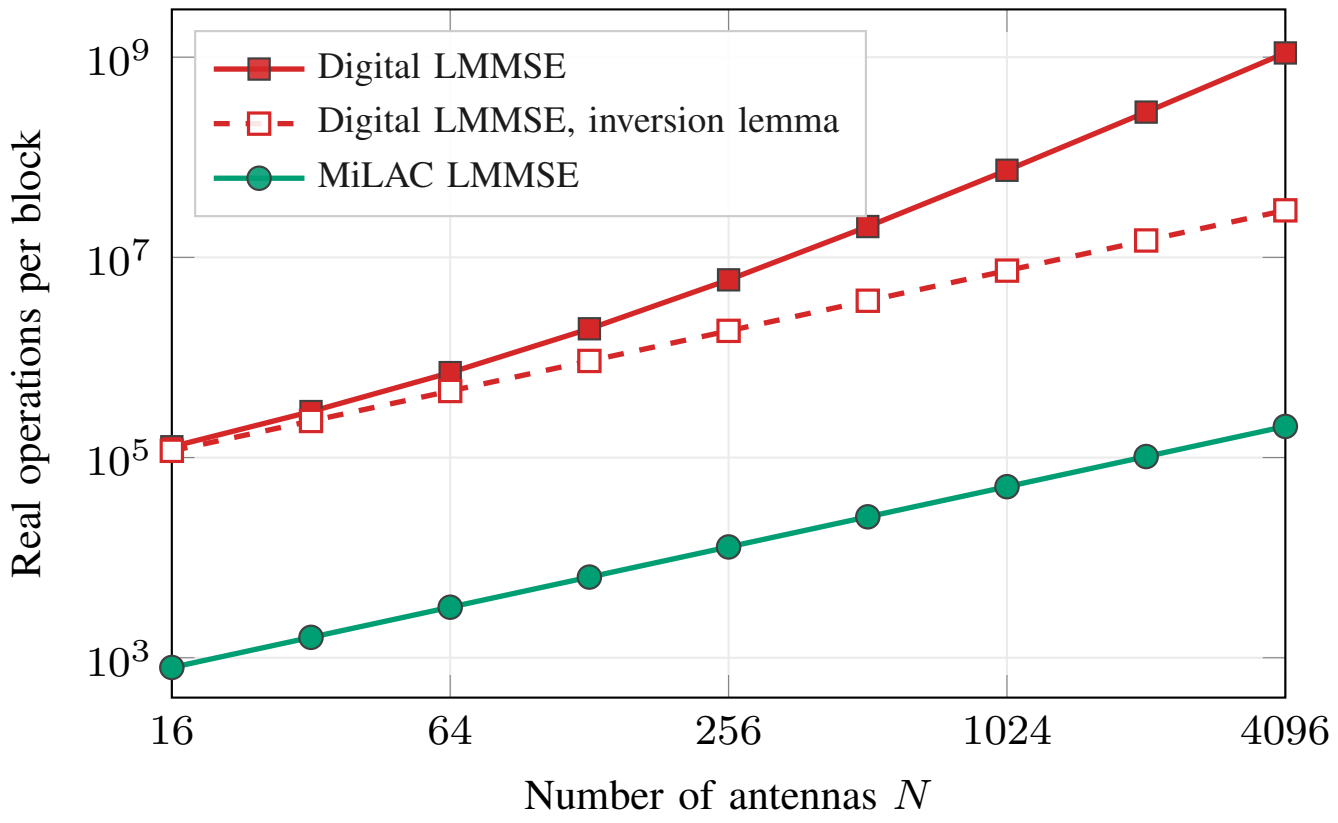
\begin{figure}[t]
\centering
\begin{tikzpicture}
\begin{axis}[
  golden axis,
  xmode=log,ymode=log,log basis x={2},log basis y={10},
  xlabel={Number of antennas $N$},ylabel={Real operations per block},
  xmin=16,xmax=4096,ymin=4e2,ymax=3e9,
  xtick={16,64,256,1024,4096},xticklabels={$16$,$64$,$256$,$1024$,$4096$},
  grid=both,grid style={gray!16},
  tick label style={font=\footnotesize},label style={font=\footnotesize},
  legend style={font=\scriptsize,at={(0.02,0.97)},anchor=north west,draw=gray!40,fill=white,fill opacity=0.9},legend cell align=left]
\addplot[curve=pltred,mark=square*,mark size=2pt] coordinates {(16,1.283e5)(32,2.881e5)(64,7.059e5)(128,1.935e6)(256,5.965e6)(512,2.032e7)(1024,7.419e7)(2048,2.826e8)(4096,1.102e9)};
\addlegendentry{Digital \ac{LMMSE}}
\addplot[curve=pltred,dashed,mark=square*,mark size=2pt,mark options={solid,fill=white,draw=pltred,line width=0.6pt}] coordinates {(16,1.168e5)(32,2.32e5)(64,4.624e5)(128,9.232e5)(256,1.845e6)(512,3.688e6)(1024,7.374e6)(2048,1.475e7)(4096,2.949e7)};
\addlegendentry{Digital \ac{LMMSE}, inversion lemma}
\addplot[curve=pltgreen,mark=*,mark size=2.2pt] coordinates {(16,800)(32,1600)(64,3200)(128,6400)(256,12800)(512,25600)(1024,51200)(2048,102400)(4096,204800)};
\addlegendentry{\ac{MiLAC} \ac{LMMSE}}
\end{axis}
\end{tikzpicture}
\caption{Computational complexity, in real operations per coherence block, of the digital and \ac{MiLAC} \ac{LMMSE} receivers against the number of antennas $N$, for $M=8$ streams, $\tau=100$ symbols per block, an isotropic prior, and an unstructured noise covariance or, for the inversion lemma, rank-$2$ interference.}
\label{fig:simcomp}
\end{figure}

For the \ac{MiLAC}, only the channel block $\Fm=g\lift{\Hm^\HH}$ is rewritten, and its susceptances are obtained from \eqref{eq:components}. Each of the $2MN$ real numbers that form $\Hm$ is scaled by $Y_0g$ once, and every inter-port susceptance of $\Fm$ is one of these values up to a sign, which costs $2MN$ multiplications. The port-to-ground susceptances are the column sums of $\Bm$ in \eqref{eq:components}. For the contribution of $\Fm$ to these sums, each column of $\Hm^\HH$, indexed by an antenna, and each row, indexed by a stream, needs the sum of the real parts and the sum of the imaginary parts of its entries. The two ports of each antenna or stream then need the sum and the difference of these two sums, so this contribution costs $2M$ additions per antenna and $2N$ per stream, i.e., $4MN$ in total. Adding the precomputed row sums of $\Dm$ at the $2N$ input ports costs $2N$ further additions, and a non-isotropic prior would add $2M$ more for the row sums of $\Em$. With a noise stage and an isotropic prior, as in Section~\ref{sec:sim-rate}, the \ac{MiLAC} \ac{LMMSE} receiver thus costs exactly $6MN+2N$ real operations per coherence block, which matches, to leading order, the approximately $6MN$ operations of the unconstrained \ac{MiLAC} \cite{Nerini2025AnalogI}. The \ac{MiLAC} \ac{RLS} receiver, which has no noise stage, costs exactly $6MN$. As discussed in Sections~\ref{sec:rzf-cost} and \ref{sec:lmmse-cost}, neither \ac{MiLAC} receiver requires per-symbol arithmetic.

Fig.~\ref{fig:simcomp} shows the number of real operations per coherence block versus the number of antennas $N$, for $M=8$ streams and $\tau=100$ symbols per block. As shown in the figure, the \ac{MiLAC} reduces this number by a factor that grows with the array size, from $1.6\times10^2$ at $N=16$ to $5.4\times10^3$ at $N=4096$. This is because the digital inverse-covariance weighting grows quadratically in $N$, whereas the number of operations of the \ac{MiLAC} grows linearly. If a digital receiver knows, as the realization does, that $\Cm_n=\sigma^2\Id+\Um\Um^\HH$ with $\Um\in\mathbb{C}^{N\times r}$, as in case (ii) of Proposition~\ref{prop:lmmse-hw}, it can instead weight the channel through the matrix inversion lemma. It computes $\Cm_n^{-1}\Hm=\sigma^{-2}(\Hm-\Um\Phim^{-1}\Um^\HH\Hm)$, with $\Phim=\sigma^2\Id_r+\Um^\HH\Um$ inverted offline. The products $\Um^\HH\Hm$ and $\Um(\Phim^{-1}\Um^\HH\Hm)$ cost $8rMN$ each, the product with $\Phim^{-1}$ costs $8r^2M$, and the subtraction and the scaling by $\sigma^{-2}$ cost $4MN$. The weighting thus costs $16rMN+8r^2M+4MN$ instead of $8MN^2$, and the other terms are unchanged, so the receiver costs $16rMN+8r^2M+4MN+8M^2N+\tfrac83M^3+8\tau MN$ per coherence block. Fig.~\ref{fig:simcomp} also shows this receiver for the rank-$2$ interference of Section~\ref{sec:sim-rate}. Its cost and that of the \ac{MiLAC} are both linear in $N$, so their ratio decreases only from $146$ at $N=16$ to $144$ at $N=4096$. The decrease comes from the terms $8r^2M$ and $\tfrac83M^3$, which do not grow with $N$, and the ratio is dominated by the digital per-symbol arithmetic $8\tau MN$.

\subsection{Circuit Complexity}
\label{sec:sim-circ}

The number of components follows from the nonzero entries of $\Bm$ through \eqref{eq:components}. As a baseline, we consider a fully-connected \ac{MiLAC} on the same $2M+2N$ ports, whose number of components is given by \eqref{eq:Nfull}. This baseline excludes the auxiliary ports, which makes the comparison conservative for \ac{LMMSE}. Moreover, comparing within the lossless and reciprocal class keeps the component type fixed, since every component is one tunable susceptance, whereas a \ac{MiLAC} realizing \eqref{eq:hermplace} also needs conductances and non-reciprocal components.

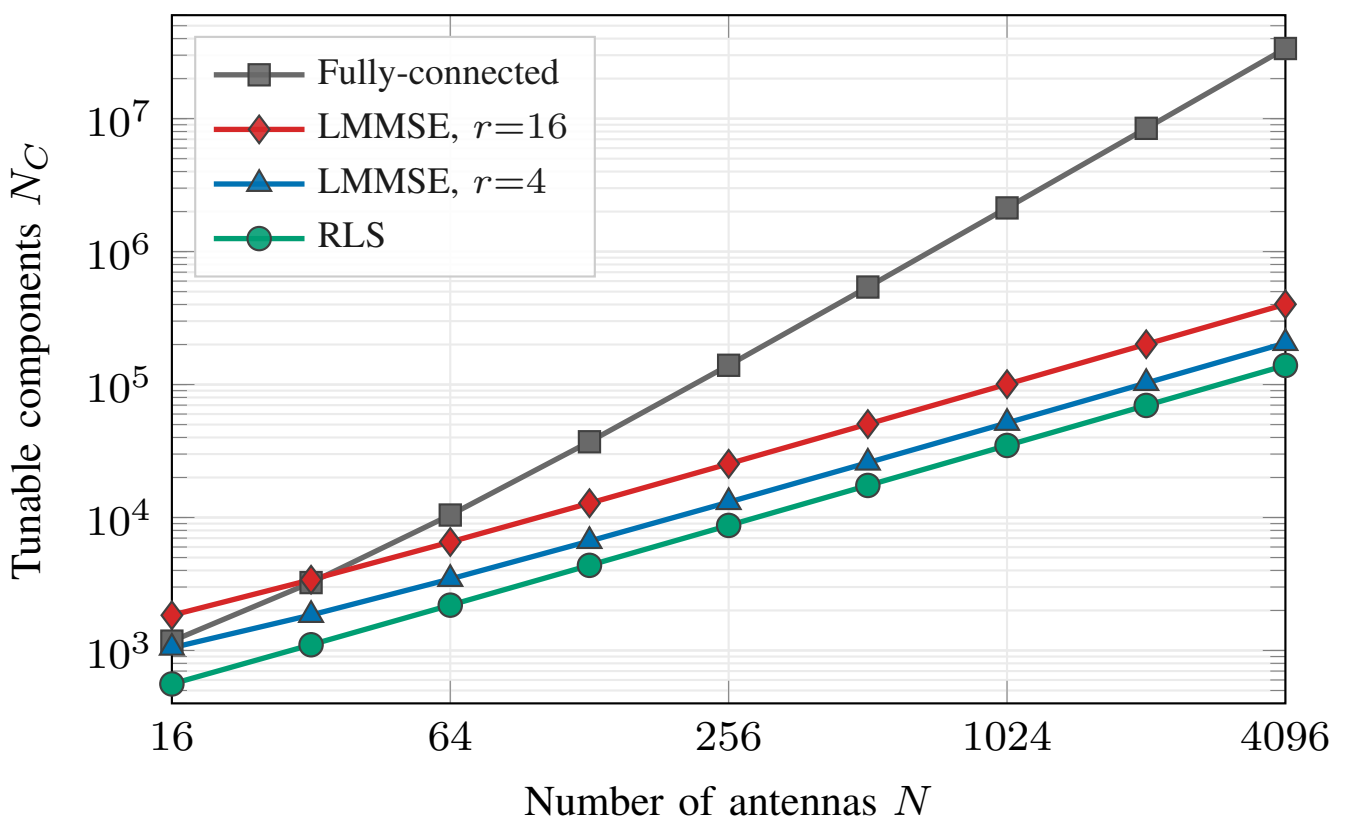
\begin{figure}[t]
\centering
\begin{tikzpicture}
\begin{axis}[
  golden axis,
  xmode=log,ymode=log,log basis x={2},log basis y={10},
  xlabel={Number of antennas $N$},ylabel={Tunable components $N_C$},
  xmin=16,xmax=4096,ymin=4e2,ymax=6e7,
  xtick={16,64,256,1024,4096},xticklabels={$16$,$64$,$256$,$1024$,$4096$},
  grid=both,grid style={gray!16},
  tick label style={font=\footnotesize},label style={font=\footnotesize},
  legend style={font=\scriptsize,at={(0.02,0.97)},anchor=north west,draw=gray!40,fill=white,fill opacity=0.9},legend cell align=left]
\addplot[curve=pltgray,mark=square*,mark size=2pt] coordinates {(16,1176)(32,3240)(64,10440)(128,37128)(256,139656)(512,541320)(1024,2131080)(2048,8456328)(4096,33689736)};
\addlegendentry{Fully-connected}
\addplot[curve=pltred,mark=diamond*,mark size=2.6pt] coordinates {(16,1840)(32,3408)(64,6544)(128,12816)(256,25360)(512,50448)(1024,100624)(2048,200976)(4096,401680)};
\addlegendentry{\ac{LMMSE}, $r{=}16$}
\addplot[curve=pltblue,mark=triangle*,mark size=2.6pt] coordinates {(16,1048)(32,1848)(64,3448)(128,6648)(256,13048)(512,25848)(1024,51448)(2048,102648)(4096,205048)};
\addlegendentry{\ac{LMMSE}, $r{=}4$}
\addplot[curve=pltgreen,mark=*,mark size=2.2pt] coordinates {(16,560)(32,1104)(64,2192)(128,4368)(256,8720)(512,17424)(1024,34832)(2048,69648)(4096,139280)};
\addlegendentry{\ac{RLS}}
\end{axis}
\end{tikzpicture}
\caption{Circuit complexity, in tunable components, of a fully-connected \ac{MiLAC} on the same $2M{+}2N$ ports and of the \ac{RLS} and \ac{LMMSE} realizations against the number of antennas $N$, for $M=8$ streams and, for \ac{LMMSE}, interference ranks $r=4$ and $16$, a non-isotropic prior, $s=1/\sigma^2$ and the minimal $\eta$ of Algorithm~\ref{alg:lmmse}, and the factors of Section~\ref{sec:hw-lmmse}. The \ac{RLS} curve also applies to \ac{LMMSE} with white noise and an isotropic prior.}
\label{fig:simhw}
\end{figure}

Fig.~\ref{fig:simhw} shows the number of tunable components versus the number of antennas $N$, for $M=8$ streams and interference ranks $r=4$ and $16$. We make three observations. First, the number of components of the fully-connected \ac{MiLAC} grows quadratically in $N$, whereas those of the proposed realizations grow linearly. At $N=4096$, the proposed realizations use $84$ times fewer components for \ac{LMMSE} with rank-$16$ interference and $242$ times fewer for \ac{RLS}. Second, the \ac{LMMSE} realizations need the components of the \ac{RLS} realization plus those of the prior stage, $O(M^2)$, and of the noise stage, $O(Nr)$. The gap to the \ac{RLS} curve is due to the interference in the noise covariance and to the non-isotropic prior, and the noise stage dominates it once $Nr\gg M^2$. Third, the saving requires a sufficiently large array: the rank-$16$ curve lies above the fully-connected one at $N=16$ and $N=32$ and drops below it only from $N=35$ on. Equating their leading terms, $2(M+N)^2$ and $4MN+4rN+4M^2$, places the crossing near $N=r+\sqrt{r^2+M^2}$, which is $34$ for $r=16$ and $13$ for $r=4$, so a higher interference rank requires a larger array.

\subsection{Performance Versus Complexity}
\label{sec:sim-rate}

We evaluate the performance on an uplink in which $M=8$ single-antenna users transmit unit-power symbols ($\Cm_x=\Id$) to a \ac{BS} with $N=256$ antennas over an \ac{i.i.d.}\ Rayleigh channel $\Hm$, so that the \ac{BS} observes
\begin{equation}
\label{eq:simmodel}
\yv=\Hm\xv+\nv,\qquad \Cm_n=\sigma^2\Id+\Um\Um^\HH,\ \ \Um\in\mathbb{C}^{N\times2},
\end{equation}
where $\nv$ is the sum of white noise of variance $\sigma^2$, with \ac{SNR} $1/\sigma^2=10$~dB, and rank-$2$ interference. The matrix $\Um$ is drawn with \ac{i.i.d.}\ circularly symmetric complex Gaussian entries and scaled so that $\|\Um\|_F^2=N/2$, which places the interference $7$~dB above the total thermal noise. The \ac{BS} applies a linear receiver $\hat\xv=\Gm\yv$ and achieves the sum rate
\begin{equation}
\label{eq:sumrate}
R=\textstyle\sum_{i=1}^{M}\log_2(1+\gamma_i),
\end{equation}
where $\gamma_i=|[\Gm\Hm]_{i,i}|^2/(\sum_{k\ne i}|[\Gm\Hm]_{i,k}|^2+[\Gm\Cm_n\Gm^\HH]_{i,i})$ is the \ac{SINR} of stream $i$ at the receiver output. The sum rate is averaged over $10^4$ independent draws of $\Hm$ and $\Um$.

\begin{figure}[t]
\centering
\begin{tikzpicture}
\begin{axis}[
  golden axis,
  xmode=log,
  xlabel={Computational complexity (real operations per block)},
  ylabel={Sum rate (bit/s/Hz)},
  xmin=8e3,xmax=1e7,ymin=35,ymax=100,
  grid=both,grid style={gray!16},
  tick label style={font=\footnotesize},label style={font=\footnotesize},
  legend style={font=\scriptsize,at={(0.40,0.03)},anchor=south,draw=gray!40,fill=white,fill opacity=0.92},legend cell align=left]
\draw[black!60,dashed,thick,{Latex[length=2.6mm,width=2mm]}-{Latex[length=2.6mm,width=2mm]},shorten <=6pt,shorten >=4.5pt] (axis cs:12800,90.1)--(axis cs:5.965e6,90.1);
\node[font=\scriptsize,fill=white,inner sep=1.5pt] at (axis cs:276322,90.1) {$466\times$ lower complexity};
\addplot[pltgreen,only marks,mark=star,mark size=5pt,mark options={line width=1.3pt}] coordinates {(12800,90.1)};
\addlegendentry{\ac{MiLAC} \ac{LMMSE}}
\addplot[curve=pltgreen,only marks,mark=pentagon*,mark size=3.6pt] coordinates {(12288,71.4)};
\addlegendentry{\ac{MiLAC} \ac{RLS}}
\addplot[curve=pltred,only marks,mark=square*,mark size=3.2pt] coordinates {(5.965e6,90.1)};
\addlegendentry{Digital \ac{LMMSE}}
\addplot[curve=pltred,only marks,mark=square*,mark size=3.2pt,mark options={solid,fill=white,draw=pltred,line width=0.8pt}] coordinates {(1.845e6,90.1)};
\addlegendentry{Digital \ac{LMMSE}, inversion lemma}
\addplot[curve=pltorange,only marks,mark=diamond*,mark size=3.8pt] coordinates {(1.771e6,71.4)};
\addlegendentry{Digital \ac{RLS}}
\addplot[curve=pltpurple,only marks,mark=triangle*,mark size=3.8pt] coordinates {(1.638e6,41.6)};
\addlegendentry{Digital \ac{MF}}
\end{axis}
\end{tikzpicture}
\caption{Sum rate versus computational complexity per coherence block of the \ac{MiLAC} \ac{LMMSE} and \ac{RLS} receivers and the digital \ac{LMMSE} (with and without the inversion lemma), \ac{RLS}, and \ac{MF} receivers, in an uplink with $M=8$ single-antenna unit-power users, $N=256$ \ac{BS} antennas, rank-$2$ interference, $10$~dB \ac{SNR}, $\lambda=\sigma^2$, and $\tau=100$ symbols per block, averaged over $10^4$ \ac{i.i.d.}\ Rayleigh channel realizations.}
\label{fig:simrate}
\end{figure}
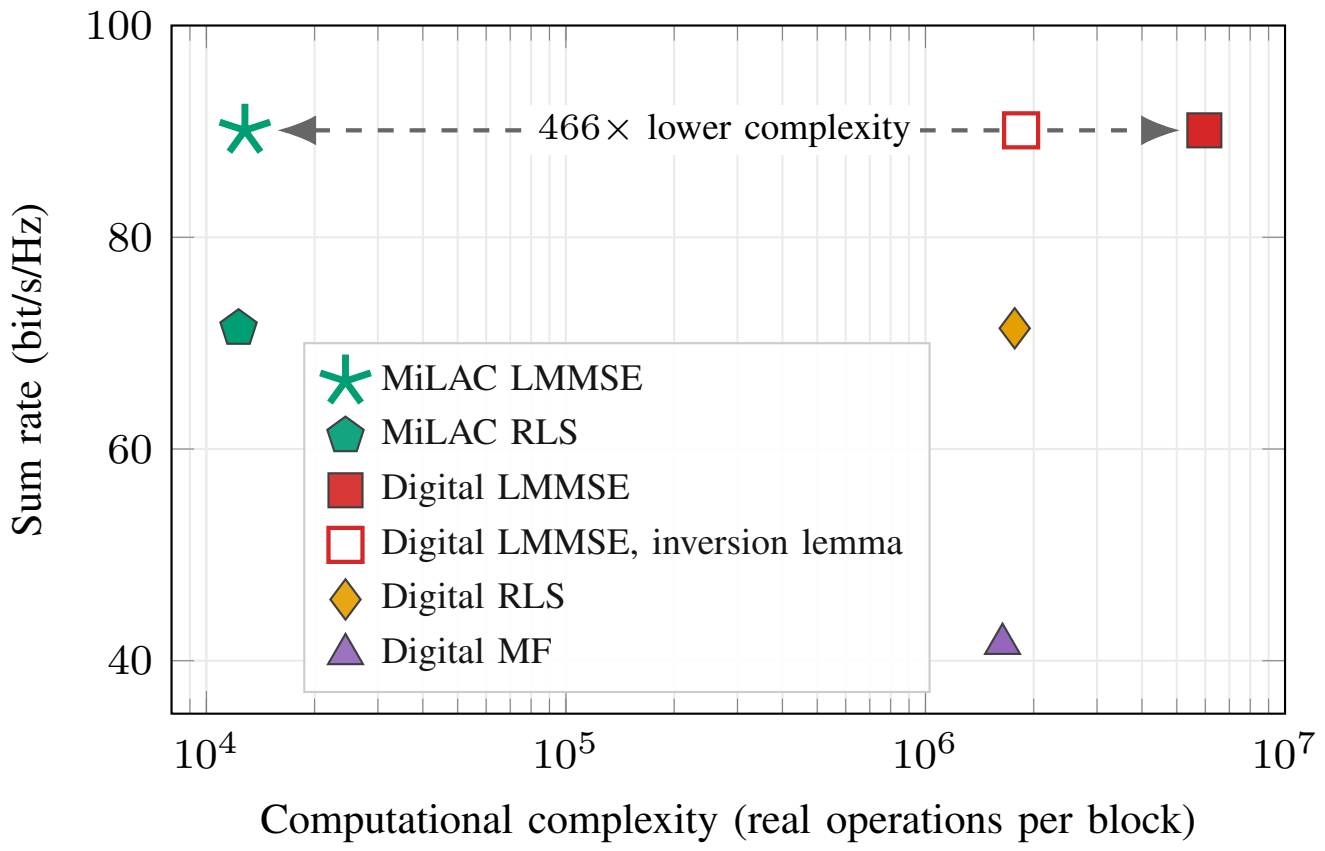

Fig.~\ref{fig:simrate} shows the sum rate versus the computational complexity per coherence block, evaluated as in Section~\ref{sec:sim-comp} with $\tau=100$, for the \ac{MiLAC} \ac{LMMSE} and \ac{RLS} receivers and the digital \ac{LMMSE}, \ac{RLS}, and \ac{MF} receivers. The \ac{RLS} receivers use $\lambda=\sigma^2$, for which \eqref{eq:rzf} equals \eqref{eq:lmmse} with $\Cm_n=\sigma^2\Id$ and $\Cm_x=\Id$. As guaranteed by Theorems~\ref{thm:rzf} and \ref{thm:lmmse}, each \ac{MiLAC} receiver attains the sum rate of its digital counterpart. The \ac{LMMSE} rate is the highest, and the \ac{MiLAC} \ac{LMMSE} receiver reaches it with $466$ times fewer operations than the digital \ac{LMMSE} receiver, as marked by the dashed arrow. This digital receiver (filled marker) treats $\Cm_n$ as unstructured, as in Fig.~\ref{fig:simcomp}. The digital receiver that weights the channel through the matrix inversion lemma (hollow marker) needs $1.84\times10^6$ operations, $144$ times more than the \ac{MiLAC} \ac{LMMSE} receiver, as in Fig.~\ref{fig:simcomp} at $N=256$. Similarly, the \ac{MiLAC} \ac{RLS} receiver needs $1.23\times10^4$ operations, against $1.77\times10^6$ for the digital \ac{RLS} receiver. The \ac{RLS} and \ac{MF} receivers achieve lower rates than \ac{LMMSE} because they ignore the interference covariance, and the \ac{MF} also leaves the inter-stream interference unsuppressed.

These results measure the cost of the computation, not the performance of a hardware implementation. They assume exactly tuned, lossless susceptances and ideal hybrid couplers, under which the realized sum rate equals the digital one because the realization is exact. Tuning resolution, component loss, and front-end insertion loss are not modeled.

\section{Conclusion}
\label{sec:conc}

In this paper, we have shown that lossless and reciprocal \acp{MiLAC} can compute the \ac{RLS} and \ac{LMMSE} estimators exactly, up to a known scaling factor, with no digital matrix inversion when the channel changes. This was an open question, since the existing \ac{MiLAC} computation of these estimators required lossy, active, or non-reciprocal components. The proposed construction doubles the input and output ports to embed the complex channel in a purely imaginary and symmetric admittance matrix, and couples terminated auxiliary ports to the input or output ports to provide the covariance weights. Based on the analysis and numerical results, we draw two conclusions. First, to leading order, the lossless and reciprocal constraint adds no computation per coherence block compared with the unconstrained \ac{MiLAC}. Indeed, when the channel changes, the realizations rewrite only the channel-dependent susceptances, at $O(MN)$ operations, against $O(MN^2)$ for a digital \ac{LMMSE} receiver with a general noise covariance and more antennas than streams. Second, the number of tunable components grows linearly with the number of antennas, for \ac{RLS} and for \ac{LMMSE} with white noise, with or without low-rank interference, rather than quadratically as for the fully-connected \ac{MiLAC} of \cite{Nerini2025AnalogI}.

Several questions remain open. The construction assumes ideal components, and the tuning resolution needed to keep the realized estimator close to its digital counterpart is left to future work. Further directions include transmit beamforming, wideband frequency-selective channels, and other linear estimators that a lossless and reciprocal \ac{MiLAC} can compute without digital matrix inversion when the channel changes.

\appendix

\subsection{Proof of Propositions~\ref{prop:class} and \ref{prop:bound}}
\label{app:class}
For Proposition~\ref{prop:class}, by \eqref{eq:admittance}, the time-averaged power entering the network through its ports is $\tfrac12\realp{\vv^\HH\iv}=\tfrac14\vv^\HH(\Ym+\Ym^\HH)\vv$, since $\vv^\HH\Ym^\HH\vv$ is the complex conjugate of $\vv^\HH\Ym\vv$. The network is lossless if and only if this power vanishes for every $\vv$, that is, $\Ym+\Ym^\HH=\Zerom$, and it is reciprocal if and only if $\Ym=\Ym^\TT$ \cite{Nerini2025Capacity,Pozar2011Microwave}. If both conditions hold, then
\begin{equation}
\label{eq:app-class}
\Ym\overset{(a)}{=}-\Ym^\HH\overset{(b)}{=}-\Ym^\star,
\end{equation}
where (a) follows from $\Ym+\Ym^\HH=\Zerom$, and (b) from $\Ym=\Ym^\TT$. Hence $\realp{\Ym}=\Zerom$, so $\Ym=j\Bm$ with $\Bm=\imagp{\Ym}$ real, and $\Ym=\Ym^\TT$ gives $\Bm=\Bm^\TT$. Conversely, $\Ym=j\Bm$ with $\Bm$ real and symmetric gives $\Ym^\HH=-j\Bm^\TT=-\Ym$ and $\Ym^\TT=\Ym$.

For Proposition~\ref{prop:bound}, we first derive \eqref{eq:WS21}, following \cite{Nerini2026Hybrid}. By \eqref{eq:waves}, $\mathbf a=\tfrac12(\vv+\iv/Y_0)$, and substituting \eqref{eq:interface} gives
\begin{equation}
\label{eq:app-a}
\mathbf a_1=\frac12(\vv_1+\uv-\vv_1)=\frac12\uv,\qquad
\mathbf a_2=\frac12(\vv_2-\vv_2)=\Zerom.
\end{equation}
Then, by substituting \eqref{eq:app-a} into $\vv=\mathbf a+\mathbf b$ and $\mathbf b=\Sm\mathbf a$, we obtain $\vv_2=\mathbf a_2+\Sm_{21}\mathbf a_1+\Sm_{22}\mathbf a_2=\tfrac12\Sm_{21}\uv$, which gives \eqref{eq:WS21} since $\vv_2=\Wm\uv$ for every $\uv$ by \eqref{eq:Wblock}. We next bound $\|\Wm\|_2$. For a lossless \ac{MiLAC}, $\Sm$ is unitary \cite{Nerini2025Capacity,Pozar2011Microwave}, so
\begin{equation}
\label{eq:app-bound}
\|\Wm\|_2\overset{(c)}{=}\frac12\|\Sm_{21}\|_2\overset{(d)}{\le}\frac12\|\Sm\|_2\overset{(e)}{=}\frac12,
\end{equation}
where (c) follows from \eqref{eq:WS21}, (d) holds since a submatrix has no larger spectral norm, and (e) follows from the unitarity of $\Sm$, as in \cite{Wu2026MiLAC}. This completes the proofs.

\subsection{Proof of Lemma~\ref{lem:hybrid}}
\label{app:hybrid}
We first compute the waves reflected from ports $3$ and $4$ of the hybrid coupler. As for $\mathbf a_1$ and $\mathbf a_2$ in \eqref{eq:app-a}, the source at port $1$ launches the incident wave $\alpha_1=y_k/2$ regardless of the load, and the terminated port $2$ has $\alpha_2=0$. By \eqref{eq:hybridS}, the reflected waves are then
\begin{equation}
\label{eq:app-beta}
\beta_3=\frac{j\alpha_1+\alpha_2}{\sqrt2}=\frac{jy_k}{2\sqrt2},\qquad
\beta_4=\frac{\alpha_1+j\alpha_2}{\sqrt2}=\frac{y_k}{2\sqrt2},
\end{equation}
which do not depend on the waves $\alpha_3$ and $\alpha_4$ incident from the network, since $[\Sm_{\mathrm{hyb}}]_{p,q}=0$ for $p,q\in\{3,4\}$. We next show that each of these ports is equivalent to a source. By \eqref{eq:waves}, the current that port $p\in\{3,4\}$ delivers to the connected network is $i_{\mathrm{out},p}=Y_0(\beta_p-\alpha_p)$, so that
\begin{equation}
\label{eq:app-thevenin}
v_p=\alpha_p+\beta_p=2\beta_p-Z_0\,i_{\mathrm{out},p},
\end{equation}
which describes a voltage source of voltage $2\beta_p$ in series with $Z_0$. Since $\beta_3$ and $\beta_4$ are fixed, the two sources are independent, and \eqref{eq:app-beta} gives their voltages $2\beta_4=y_k/\sqrt2$ and $2\beta_3=jy_k/\sqrt2$. This completes the proof.

\subsection{Properties of the Real-Valued Representation}
\label{app:lift}
We collect here the properties of the real-valued representation $\lift{\cdot}$ used in this paper, each followed by its derivation. For a complex matrix $\Am\in\mathbb{C}^{m\times n}$, written as $\Am=\Am_{\mathrm r}+j\Am_{\mathrm i}$ with $\Am_{\mathrm r},\Am_{\mathrm i}\in\mathbb{R}^{m\times n}$, the representation defined in \eqref{eq:lift} is
\begin{equation}
\label{eq:app-lift}
\lift{\Am}=\begin{bmatrix}\Am_{\mathrm r}&\Am_{\mathrm i}\\-\Am_{\mathrm i}&\Am_{\mathrm r}\end{bmatrix}\in\mathbb{R}^{2m\times2n}.
\end{equation}
Throughout, $\Qm=\Qm_{\mathrm r}+j\Qm_{\mathrm i}$ is a second complex matrix, with $\Qm_{\mathrm r}$ and $\Qm_{\mathrm i}$ real, and the dimensions of $\Am$ and $\Qm$ are such that the stated operations are defined.
\begin{enumerate}[label=(P\arabic*),ref=(P\arabic*),leftmargin=2.5em,itemsep=2pt]
\item\label{P:lin} \emph{Real linearity:} For real scalars $c_1$ and $c_2$, $\lift{c_1\Am+c_2\Qm}=c_1\lift{\Am}+c_2\lift{\Qm}$, and $\lift{\Id_n}=\Id_{2n}$. Both identities follow from \eqref{eq:app-lift}, since the real and imaginary parts are real linear, and those of $\Id_n$ are $\Id_n$ and $\Zerom$.
\item\label{P:prod} \emph{Products:} The representation of a product is the product of the representations, $\lift{\Am\Qm}=\lift{\Am}\lift{\Qm}$. Multiplying the blocks of \eqref{eq:app-lift} gives a matrix whose blocks are $\realp{\Am\Qm}=\Am_{\mathrm r}\Qm_{\mathrm r}-\Am_{\mathrm i}\Qm_{\mathrm i}$ and $\imagp{\Am\Qm}=\Am_{\mathrm r}\Qm_{\mathrm i}+\Am_{\mathrm i}\Qm_{\mathrm r}$, arranged as in \eqref{eq:app-lift}.
\item\label{P:inv} \emph{Inverses:} If $\Am$ is square and invertible, then $\lift{\Am}$ is invertible and $\lift{\Am^{-1}}=\lift{\Am}^{-1}$. By \ref{P:prod} and \ref{P:lin}, $\lift{\Am}\lift{\Am^{-1}}=\lift{\Am\Am^{-1}}=\lift{\Id}=\Id$.
\item\label{P:herm} \emph{Conjugate transposes:} The representation of $\Am^\HH$ is the transpose of that of $\Am$, $\lift{\Am^\HH}=\lift{\Am}^\TT$, so $\lift{\Am}$ is symmetric if $\Am$ is Hermitian. This follows since $\realp{\Am^\HH}=\Am_{\mathrm r}^\TT$ and $\imagp{\Am^\HH}=-\Am_{\mathrm i}^\TT$, and \eqref{eq:app-lift} applied to $\Am^\HH$ gives $\lift{\Am}^\TT$.
\item\label{P:pair} \emph{Signal pairs:} For every $\zv\in\mathbb{C}^n$, $\lift{\Am}[\zv^\TT,\,j\zv^\TT]^\TT=[(\Am\zv)^\TT,\,j(\Am\zv)^\TT]^\TT$. By \eqref{eq:app-lift}, the first block of the product is $\Am_{\mathrm r}\zv+\Am_{\mathrm i}(j\zv)=\Am\zv$, and the second is $-\Am_{\mathrm i}\zv+\Am_{\mathrm r}(j\zv)=j(\Am_{\mathrm r}+j\Am_{\mathrm i})\zv=j\Am\zv$.
\item\label{P:spec} \emph{Spectrum:} If $\Am\in\mathbb{C}^{n\times n}$ is Hermitian, the eigenvalues of $\lift{\Am}$ are those of $\Am$, each with doubled multiplicity, so $\lift{\Am}$ is positive semidefinite (definite) if and only if $\Am$ is. To show this, let $\Am=\Vm\,\diag(\mu_1,\dots,\mu_n)\Vm^\HH$, where $\Vm$ is unitary and the diagonal matrix $\diag(\mu_1,\dots,\mu_n)$ contains the real eigenvalues of $\Am$. Then
\begin{align*}
\lift{\Am}&\overset{(a)}{=}\lift{\Vm}\,\lift{\diag(\mu_1,\dots,\mu_n)}\,\lift{\Vm}^\TT\\
&\overset{(b)}{=}\lift{\Vm}\,\diag(\mu_1,\dots,\mu_n,\mu_1,\dots,\mu_n)\,\lift{\Vm}^\TT,
\end{align*}
where (a) follows from \ref{P:prod} and \ref{P:herm}, and (b) from \eqref{eq:app-lift}. Since $\lift{\Vm}^\TT\lift{\Vm}=\lift{\Vm^\HH\Vm}=\Id$ by \ref{P:herm}, \ref{P:prod}, and \ref{P:lin}, this is an eigenvalue decomposition of $\lift{\Am}$, which proves the statement.
\item\label{P:norm} \emph{Norm and rank:} The representation preserves the spectral norm and doubles the rank, $\|\lift{\Am}\|_2=\|\Am\|_2$ and $\rank(\lift{\Am})=2\rank(\Am)$. By \ref{P:prod} and \ref{P:herm}, $\lift{\Am}^\TT\lift{\Am}=\lift{\Am^\HH\Am}$, which by \ref{P:spec} has the eigenvalues of the Hermitian matrix $\Am^\HH\Am$, each with doubled multiplicity. Hence $\|\lift{\Am}\|_2^2=\lmax(\Am^\HH\Am)=\|\Am\|_2^2$, and $\rank(\lift{\Am})=\rank(\lift{\Am}^\TT\lift{\Am})=2\rank(\Am^\HH\Am)=2\rank(\Am)$, since the rank of a Hermitian matrix is the number of its nonzero eigenvalues.
\end{enumerate}

\subsection{Proof of Theorem~\ref{thm:rzf} and Corollary~\ref{cor:limits}}
\label{app:rzf}
The matrix \eqref{eq:Brzf} is real and symmetric, so the \ac{MiLAC} is lossless and reciprocal by Proposition~\ref{prop:class}. We first derive \eqref{eq:rzfmap} and \eqref{eq:rzfout}. Substituting \eqref{eq:Brzf} into $\Pm=j\Bm/Y_0+\Id$ gives
\begin{equation}
\label{eq:app-Przf}
\Pm=\begin{bmatrix}\Id & j\Fm^\TT\\ j\Fm & \Id\end{bmatrix},
\end{equation}
in which the Schur complement of the input block is $\Id-(j\Fm)(j\Fm^\TT)=\Id+\Fm\Fm^\TT$, a positive definite matrix. Block matrix inversion then gives
\begin{align}
\bigl[\Pm^{-1}\bigr]_{21}&=-(\Id+\Fm\Fm^\TT)^{-1}(j\Fm)\nonumber\\
&\overset{(a)}{=}-j\,\Bigl(\frac1\lambda\lift{\lambda\Id+\Hm^\HH\Hm}\Bigr)^{-1}\frac{1}{\sqrt\lambda}\lift{\Hm^\HH}\nonumber\\
&\overset{(b)}{=}-j\sqrt\lambda\,\lift{(\lambda\Id+\Hm^\HH\Hm)^{-1}\Hm^\HH}\nonumber\\
&=-j\sqrt\lambda\,\lift{\Wm_{\mathrm{RLS}}},\label{eq:app-Wrzf}
\end{align}
where (a) follows from $\Id+\Fm\Fm^\TT=\Id+\lift{\Hm^\HH}\lift{\Hm}/\lambda=\lift{\lambda\Id+\Hm^\HH\Hm}/\lambda$, which holds by \eqref{eq:Brzf}, \ref{P:herm}, \ref{P:prod}, and \ref{P:lin}, and (b) follows from \ref{P:inv} and \ref{P:prod}. Then, substituting \eqref{eq:app-Wrzf} and the source vector \eqref{eq:quadsource} of Lemma~\ref{lem:hybrid} into $\vv_2=[\Pm^{-1}]_{21}\uv$ gives $\vv_2=-j\sqrt{\lambda/2}\,\lift{\Wm_{\mathrm{RLS}}}[\yv^\TT,j\yv^\TT]^\TT$, which is \eqref{eq:rzfout} by \ref{P:pair}.

We next prove the norm bound. Let $\nu_1,\dots,\nu_{\min(M,N)}$ be the singular values of $\Hm$. By the singular value decomposition of $\Hm$, the singular values of $\sqrt\lambda\,\Wm_{\mathrm{RLS}}$ are
\begin{equation}
\label{eq:app-sv}
\frac{\sqrt\lambda\,\nu_i}{\lambda+\nu_i^2}\overset{(c)}{\le}\frac12,
\end{equation}
where (c) follows from $\lambda+\nu_i^2-2\sqrt\lambda\,\nu_i=(\sqrt\lambda-\nu_i)^2\ge0$, with equality if and only if $\lambda=\nu_i^2$. By \ref{P:norm}, the realized block \eqref{eq:rzfmap} thus has spectral norm $\sqrt\lambda\,\|\Wm_{\mathrm{RLS}}\|_2\le\tfrac12$, with equality if and only if $\lambda=\nu_i^2$ for some $i$.

Finally, for Corollary~\ref{cor:limits}, Theorem~\ref{thm:rzf} holds for every $\lambda>0$. If $\Hm$ has full column rank, $\Hm^\HH\Hm$ is invertible, and the continuity of the matrix inverse gives $\Wm_{\mathrm{RLS}}\to(\Hm^\HH\Hm)^{-1}\Hm^\HH$ as $\lambda\to0$. For every $\Hm$, the continuity of the matrix inverse at $\Id$ also gives $\lambda\Wm_{\mathrm{RLS}}=(\Hm^\HH\Hm/\lambda+\Id)^{-1}\Hm^\HH\to\Hm^\HH$ as $\lambda\to\infty$. This completes the proofs.

\subsection{Proof of Theorem~\ref{thm:lmmse}}
\label{app:gen}
The matrix \eqref{eq:Btemplate} is real and symmetric, so the \ac{MiLAC} is lossless and reciprocal by Proposition~\ref{prop:class}, and the factors $\Dm$ and $\Em$ exist under the conditions on $s$ and $\eta$, as shown before Theorem~\ref{thm:lmmse}. The proof has three steps: we eliminate the auxiliary ports, invert the reduced matrix \eqref{eq:Peff}, and identify the result with $\Wm_{\mathrm{LMMSE}}$.

\emph{Step 1 (elimination of the auxiliary ports):} Denote by $\mathcal K$ the set of the $2N$ input and $2M$ output ports and by $\mathcal A$ the set of the $m_{\mathrm D}+m_{\mathrm E}$ auxiliary ports. With the ports of $\mathcal K$ placed first, we partition $\Pm=j\Bm/Y_0+\Id$, the port voltages, and the source vector of \eqref{eq:iomap} as
\begin{equation}
\label{eq:partition}
\Pm=\begin{bmatrix}\Pm_{\mathcal{KK}}&\Pm_{\mathcal{KA}}\\ \Pm_{\mathcal{AK}}&\Pm_{\mathcal{AA}}\end{bmatrix},\qquad
\vv=\begin{bmatrix}\vv_{\mathcal K}\\ \vv_{\mathcal A}\end{bmatrix},\qquad
\tilde\uv=\begin{bmatrix}\tilde\uv_{\mathcal K}\\ \Zerom\end{bmatrix},
\end{equation}
where the source vector vanishes at the auxiliary ports because no source is connected to them. With the ports of $\mathcal K$ ordered as input and then as output ports, \eqref{eq:Btemplate} gives
\begin{equation}
\label{eq:app-blocks}
\Pm_{\mathcal{KK}}=\begin{bmatrix}\Id & j\Fm^\TT\\ j\Fm & \Id\end{bmatrix},\qquad
\Pm_{\mathcal{KA}}=\Pm_{\mathcal{AK}}^\TT=j\begin{bmatrix}\Dm & \Zerom\\ \Zerom & \Em\end{bmatrix},
\end{equation}
and $\Pm_{\mathcal{AA}}=\Id$, the identity contributed by the terminations, since the auxiliary ports are not coupled to each other and the diagonal entries of $\Bm$ at these ports are zero. The second block row of $\Pm\vv=\tilde\uv$ thus gives $\vv_{\mathcal A}=-\Pm_{\mathcal{AK}}\vv_{\mathcal K}$, and by substituting it into the first block row, $\Pm_{\mathcal{KK}}\vv_{\mathcal K}+\Pm_{\mathcal{KA}}\vv_{\mathcal A}=\tilde\uv_{\mathcal K}$, we obtain
\begin{equation}
\label{eq:schur}
\bigl(\Pm_{\mathcal{KK}}-\Pm_{\mathcal{KA}}\Pm_{\mathcal{AK}}\bigr)\vv_{\mathcal K}=\tilde\uv_{\mathcal K},
\end{equation}
where, by \eqref{eq:app-blocks}, $-\Pm_{\mathcal{KA}}\Pm_{\mathcal{AK}}$ is block diagonal with the blocks $\Dm\Dm^\TT$ and $\Em\Em^\TT$. By \eqref{eq:DEFlmmse} and \eqref{eq:Elmmse}, the matrix in \eqref{eq:schur} is thus $\Pm_{\mathrm{eff}}$ in \eqref{eq:Peff}, which Step~2 shows to be invertible. Since $\tilde\uv_{\mathcal K}=[\uv^\TT,\Zerom^\TT]^\TT$, the voltages at the output ports are $\vv_{\mathrm o}=[\Pm_{\mathrm{eff}}^{-1}]_{21}\uv$ for every $\uv$, so $[\Pm^{-1}]_{\mathrm{oi}}=[\Pm_{\mathrm{eff}}^{-1}]_{21}$.

\emph{Step 2 (block matrix inversion):} The input block $s\lift{\Cm_n}$ of \eqref{eq:Peff} is positive definite by \ref{P:spec}, and its Schur complement is
\begin{align}
\Gammam_{\mathrm{o}}&=\eta\lift{\Cm_x^{-1}}-(j\Fm)\bigl(s\lift{\Cm_n}\bigr)^{-1}(j\Fm^\TT)\nonumber\\
&\overset{(a)}{=}\eta\lift{\Cm_x^{-1}}+\frac{g^2}{s}\,\lift{\Hm^\HH\Cm_n^{-1}\Hm}\nonumber\\
&\overset{(b)}{=}\eta\,\lift{\Hm^\HH\Cm_n^{-1}\Hm+\Cm_x^{-1}},\label{eq:app-Schur}
\end{align}
where (a) follows from $j\cdot j=-1$ and $\Fm(s\lift{\Cm_n})^{-1}\Fm^\TT=(g^2/s)\lift{\Hm^\HH}\lift{\Cm_n^{-1}}\lift{\Hm}=(g^2/s)\lift{\Hm^\HH\Cm_n^{-1}\Hm}$, which holds by \eqref{eq:Flmmse}, \ref{P:inv}, \ref{P:herm}, and \ref{P:prod}, and (b) follows from $g^2=s\eta$ and \ref{P:lin}. By \ref{P:spec}, $\Gammam_{\mathrm{o}}$ is positive definite, so $\Pm_{\mathrm{eff}}$ is invertible, and block matrix inversion gives $[\Pm_{\mathrm{eff}}^{-1}]_{21}=-\Gammam_{\mathrm{o}}^{-1}(j\Fm)(s\lift{\Cm_n})^{-1}$.

\emph{Step 3 (identification):} By substituting \eqref{eq:app-Schur} and \eqref{eq:Flmmse} into this expression, we obtain
\begin{align}
\bigl[\Pm_{\mathrm{eff}}^{-1}\bigr]_{21}&\overset{(c)}{=}-\frac{jg}{s\eta}\,\lift{(\Hm^\HH\Cm_n^{-1}\Hm+\Cm_x^{-1})^{-1}}\lift{\Hm^\HH}\lift{\Cm_n^{-1}}\nonumber\\
&\overset{(d)}{=}-\frac{j}{\sqrt{s\eta}}\,\lift{\Wm_{\mathrm{LMMSE}}},\label{eq:app-Wlmmse}
\end{align}
where (c) follows from \ref{P:inv}, and (d) from \ref{P:prod} and $g/(s\eta)=1/\sqrt{s\eta}$. This proves \eqref{eq:lmmsemap}. Finally, substituting \eqref{eq:app-Wlmmse} and the source vector \eqref{eq:quadsource} of Lemma~\ref{lem:hybrid} into $\vv_{\mathrm o}=[\Pm_{\mathrm{eff}}^{-1}]_{21}\uv$ gives \eqref{eq:lmmseout} by \ref{P:pair}. This completes the proof.

\subsection{Proof of Propositions~\ref{prop:rzf-hw} and \ref{prop:lmmse-hw}}
\label{app:hw}
For Proposition~\ref{prop:rzf-hw}, \eqref{eq:Brzf} has no auxiliary ports, and its only nonzero blocks are $\Fm$ and $\Fm^\TT$, which contain the same $4MN$ entries and couple input ports only to output ports. By \eqref{eq:components}, each nonzero entry of $\Fm$ defines one inter-port susceptance, and each port with a nonzero column sum of $\Bm$ one port-to-ground susceptance. For a generic $\Hm$, all these quantities are nonzero, so the graph is complete bipartite, with $N_E=4MN$ edges and $N_V=2M+2N$ ports, and \eqref{eq:Ncount} gives \eqref{eq:Nrzf}. For other channels, an entry of $\Hm$ with a zero real or imaginary part, or a column of $\Bm$ with zero sum, only removes components, so \eqref{eq:Nrzf} is an upper bound for every $\Hm$.

For Proposition~\ref{prop:lmmse-hw}, we first show that any real factor $\Dm$ satisfying \eqref{eq:DEFlmmse} realizes the noise stage as stated. Theorem~\ref{thm:lmmse} holds for every such $\Dm$, and, in \eqref{eq:Btemplate}, each column of $\Dm$ defines one noise-stage port and, by \eqref{eq:components}, each nonzero entry of $\Dm$ one inter-port susceptance. We next derive the three cases. For $\Dm=\lift{\Dm_{\mathrm c}}$ with $\Dm_{\mathrm c}\Dm_{\mathrm c}^\HH=s\Cm_n-\Id$, we have
\begin{equation}
\label{eq:app-DDT}
\Dm\Dm^\TT\overset{(a)}{=}\lift{\Dm_{\mathrm c}\Dm_{\mathrm c}^\HH}=\lift{s\Cm_n-\Id}\overset{(b)}{=}s\lift{\Cm_n}-\Id,
\end{equation}
where (a) follows from \ref{P:herm} and \ref{P:prod}, and (b) from \ref{P:lin}, so $\Dm$ satisfies \eqref{eq:DEFlmmse} with twice as many columns as $\Dm_{\mathrm c}$. By its eigenvalue decomposition, a positive semidefinite matrix of rank $\rho$ has a factor with $\rho$ columns, and $s\Cm_n-\Id$ has rank $0$, at most $r$, and at most $N-1$ in cases (i)--(iii), the last because $s=1/\lmin(\Cm_n)$ makes it singular. Taking $\Dm_{\mathrm c}$ empty, $\Dm_{\mathrm c}=\Um/\sigma$, and a factor with $\rank(s\Cm_n-\Id)$ columns, respectively, gives $m_{\mathrm D}=0$, $2r$, and at most $2N-2$. Finally, each auxiliary port adds one port-to-ground susceptance and at most one inter-port susceptance per row of its factor. By \eqref{eq:Ncount}, the number of components is thus at most $4MN+2M+2N$, as in Proposition~\ref{prop:rzf-hw}, plus $m_{\mathrm E}(2M+1)\le2M(2M+1)$ for the prior stage and $m_{\mathrm D}(2N+1)$ for the noise stage. For $M\le N$, the three values of $m_{\mathrm D}$ give the stated numbers of components. This completes the proofs.

\bibliographystyle{IEEEtran}
\bibliography{references}

@article{Nerini2025AnalogI,
  author  = {M. Nerini and B. Clerckx},
  title   = {Analog Computing for Signal Processing and Communications -- {Part I}: Computing With Microwave Networks},
  journal = {IEEE Trans. Signal Process.},
  volume  = {73},
  pages   = {5183--5197},
  year    = {2025}
}

@article{Nerini2025AnalogII,
  author  = {M. Nerini and B. Clerckx},
  title   = {Analog Computing for Signal Processing and Communications -- {Part II}: Toward Gigantic {MIMO} Beamforming},
  journal = {IEEE Trans. Signal Process.},
  volume  = {73},
  pages   = {5198--5212},
  year    = {2025}
}

@article{Wei2026Gigantic,
  author  = {A. Wei and Z. Wang and W. Du and S. Wang and D. Zhang and K. Zhi and Q. Li and B. Clerckx},
  title   = {Microwave Linear Analog Computing for Gigantic {MIMO} Systems: Fundamental Architectures, Hardware Constraints, and Adaptive Calibration},
  journal = {IEEE Wireless Commun.},
  year    = {2026},
  note    = {early access, doi: 10.1109/MWC.2026.3738746}
}

@article{Nerini2025Capacity,
  author  = {M. Nerini and B. Clerckx},
  title   = {Capacity of {MIMO} Systems Aided by Microwave Linear Analog Computers ({MiLACs})},
  journal = {arXiv:2506.05983},
  year    = {2025}
}

@article{Nerini2026Stem,
  author  = {M. Nerini and B. Clerckx},
  title   = {{MIMO} Systems Aided by Microwave Linear Analog Computers: Capacity-Achieving Architectures With Reduced Circuit Complexity},
  journal = {IEEE Trans. Wireless Commun.},
  volume  = {25},
  pages   = {14597--14610},
  year    = {2026}
}

@article{Wu2026MiLAC,
  author  = {Z. Wu and M. Nerini and B. Clerckx},
  title   = {Microwave Linear Analog Computer ({MiLAC})-Aided Multiuser {MISO}: Fundamental Limits and Beamforming Design},
  journal = {arXiv:2601.10060},
  year    = {2026}
}

@article{Fang2026Performance,
  author  = {T. Fang and X. Zhou and Y. Mao},
  title   = {On the Performance of Lossless Reciprocal {MiLAC} Architectures in Multi-User Networks},
  journal = {IEEE Wireless Commun. Lett.},
  volume  = {15},
  pages   = {2609--2613},
  year    = {2026}
}

@article{Zhou2026TwoLayer,
  author  = {X. Zhou and T. Fang and Y. Mao and B. Clerckx},
  title   = {Two-Layer Microwave Linear Analog Computer ({MiLAC})-Aided Multi-User {MISO} Networks},
  journal = {arXiv:2604.24303},
  year    = {2026}
}

@article{Zhang2026Beamforming,
  author  = {Y. Zhang and Z. Wu and B. Clerckx and T. Y. Al-Naffouri},
  title   = {Beamforming Design for Stem-Connected Microwave Linear Analog Computer ({MiLAC})-Aided Multiuser {MISO} Downlinks},
  journal = {arXiv:2606.14499},
  year    = {2026}
}

@article{Zhang2026EE,
  author  = {Y. Zhang and P. Zheng and T. Y. Al-Naffouri},
  title   = {Quantization-Aware {EE} Optimization and {SE}-{EE} Tradeoff for {MiLAC}-Aided {MU-MISO} Beamforming},
  journal = {IEEE Trans. Signal Process.},
  volume  = {74},
  pages   = {4225--4240},
  year    = {2026}
}

@article{Zhang2026ChannelEst,
  author  = {Q. Zhang and M. Nerini and B. Clerckx},
  title   = {Channel Estimation in {MIMO} Systems Aided by Microwave Linear Analog Computers ({MiLACs})},
  journal = {IEEE Wireless Commun. Lett.},
  year    = {2026},
  note    = {early access, doi: 10.1109/LWC.2026.3736221}
}

@article{Zhang2026ChannelBF,
  author  = {Q. Zhang and M. Nerini and B. Clerckx},
  title   = {Channel Estimation and Beamforming for Microwave Linear Analog Computers ({MiLACs})-Aided Multiuser {MISO} Systems},
  journal = {arXiv:2607.00954},
  year    = {2026}
}

@article{Zhang2026DOA,
  author  = {Y. Zhang and Y. Ge and B. Clerckx and T. Y. Al-Naffouri},
  title   = {How Many {RF} Chains Does a Microwave Linear Analog Computer ({MiLAC}) Need to Match the Fully-Digital {Cram\'er--Rao} Bound?},
  journal = {arXiv:2606.23986},
  year    = {2026}
}

@article{Wang2026AirComp,
  author  = {Y. Wang and D. Zhang and Q. Li and J. Liu and C. Shi},
  title   = {{MiLAC}-Aided Beamforming for {MIMO} Over-the-Air Computation},
  journal = {arXiv:2608.13353},
  year    = {2026}
}

@article{Zhang2026ISAC,
  author  = {Y. Zhang and B. Clerckx and T. Y. Al-Naffouri},
  title   = {Microwave Linear Analog Computer ({MiLAC})-Aided Integrated Sensing and Communication: Transceiver Design and Beamforming Optimization},
  journal = {arXiv preprint},
  year    = {2026}
}

@article{Nerini2026Hybrid,
  author  = {M. Nerini and X. Liu and B. Clerckx},
  title   = {Analog Computing With Hybrid Couplers and Phase Shifters},
  journal = {IEEE Trans. Microw. Theory Techn.},
  volume  = {74},
  number  = {9},
  pages   = {8039--8061},
  month   = sep,
  year    = {2026}
}

@article{Nerini2026Physics,
  author  = {M. Nerini and B. Clerckx},
  title   = {Physics-Compliant Modeling and Optimization of {MIMO} Systems Aided by Microwave Linear Analog Computers},
  journal = {arXiv:2602.19379},
  year    = {2026}
}

@article{Nerini2026ActivePassive,
  author  = {M. Nerini and B. Clerckx},
  title   = {Microwave Linear Analog Computer ({MiLAC}) for Simultaneous Active and Passive Beamforming},
  journal = {IEEE Commun. Lett.},
  volume  = {30},
  pages   = {3352--3356},
  year    = {2026}
}

@article{Zhou2026LossyEE,
  author  = {B. Zhou and B. Clerckx},
  title   = {Lossy Microwave Linear Analog Computer ({MiLAC}) for Future {MIMO}: Learning-Based Architecture Designs for Spectral and Energy Efficiency Maximization},
  journal = {arXiv:2606.02369},
  year    = {2026}
}

@article{Liu2026Radar,
  author  = {Z. Liu and Z. Wu and B. Clerckx},
  title   = {Microwave Linear Analog Computer ({MiLAC})-Aided {MIMO} Radar Sensing: Transmit Beamforming Design and {DoA} Estimation},
  journal = {arXiv:2605.21020},
  year    = {2026}
}

@article{Caloz2013Analog,
  author  = {C. Caloz and S. Gupta and Q. Zhang and B. Nikfal},
  title   = {Analog Signal Processing: A Possible Alternative or Complement to Dominantly Digital Radio Schemes},
  journal = {IEEE Microw. Mag.},
  volume  = {14},
  number  = {6},
  pages   = {87--103},
  year    = {2013}
}

@article{Solli2015Optical,
  author  = {D. R. Solli and B. Jalali},
  title   = {Analog Optical Computing},
  journal = {Nature Photon.},
  volume  = {9},
  number  = {11},
  pages   = {704--706},
  year    = {2015}
}

@article{Wu2014Optical,
  author  = {K. Wu and C. Soci and P. P. Shum and N. I. Zheludev},
  title   = {Computing Matrix Inversion With Optical Networks},
  journal = {Opt. Express},
  volume  = {22},
  number  = {1},
  pages   = {295--304},
  year    = {2014}
}

@article{Silva2014Performing,
  author  = {A. Silva and F. Monticone and G. Castaldi and V. Galdi and A. Al\`{u} and N. Engheta},
  title   = {Performing Mathematical Operations With Metamaterials},
  journal = {Science},
  volume  = {343},
  number  = {6167},
  pages   = {160--163},
  year    = {2014}
}

@article{Estakhri2019Inverse,
  author  = {N. {Mohammadi Estakhri} and B. Edwards and N. Engheta},
  title   = {Inverse-Designed Metastructures That Solve Equations},
  journal = {Science},
  volume  = {363},
  number  = {6433},
  pages   = {1333--1338},
  year    = {2019}
}

@article{Sun2019Solving,
  author  = {Z. Sun and G. Pedretti and E. Ambrosi and A. Bricalli and W. Wang and D. Ielmini},
  title   = {Solving Matrix Equations in One Step With Cross-Point Resistive Arrays},
  journal = {Proc. Nat. Acad. Sci. USA},
  volume  = {116},
  number  = {10},
  pages   = {4123--4128},
  year    = {2019}
}

@article{Wang2023Parallel,
  author  = {C. Wang and others},
  title   = {Parallel In-Memory Wireless Computing},
  journal = {Nature Electron.},
  volume  = {6},
  number  = {5},
  pages   = {381--389},
  year    = {2023}
}

@article{Bjornson2024Gigantic,
  author  = {E. Bj\"{o}rnson and F. Kara and N. Kolomvakis and A. Kosasih and P. Ramezani and M. B. Salman},
  title   = {Enabling 6{G} Performance in the Upper Mid-Band by Transitioning From Massive to Gigantic {MIMO}},
  journal = {IEEE Open J. Commun. Soc.},
  volume  = {6},
  pages   = {5450--5463},
  year    = {2025}
}

@article{Sohrabi2016Hybrid,
  author  = {F. Sohrabi and W. Yu},
  title   = {Hybrid Digital and Analog Beamforming Design for Large-Scale Antenna Arrays},
  journal = {IEEE J. Sel. Topics Signal Process.},
  volume  = {10},
  number  = {3},
  pages   = {501--513},
  year    = {2016}
}

@article{Marzetta2010Noncooperative,
  author  = {T. L. Marzetta},
  title   = {Noncooperative Cellular Wireless With Unlimited Numbers of Base Station Antennas},
  journal = {IEEE Trans. Wireless Commun.},
  volume  = {9},
  number  = {11},
  pages   = {3590--3600},
  year    = {2010}
}

@article{Peel2005Vector,
  author  = {C. B. Peel and B. M. Hochwald and A. L. Swindlehurst},
  title   = {A Vector-Perturbation Technique for Near-Capacity Multiantenna Multiuser Communication---{Part I}: Channel Inversion and Regularization},
  journal = {IEEE Trans. Commun.},
  volume  = {53},
  number  = {1},
  pages   = {195--202},
  year    = {2005}
}

@book{Kay1993Fundamentals,
  author    = {S. M. Kay},
  title     = {Fundamentals of Statistical Signal Processing: Estimation Theory},
  publisher = {Prentice-Hall},
  address   = {Englewood Cliffs, NJ, USA},
  year      = {1993}
}

@book{Pozar2011Microwave,
  author    = {D. M. Pozar},
  title     = {Microwave Engineering},
  edition   = {4th},
  publisher = {Wiley},
  address   = {Hoboken, NJ, USA},
  year      = {2012}
}

\end{document}